\ifdefined\pdfminorversion
\fi
\documentclass[a4paper,11pt]{article}

\usepackage[T1]{fontenc}
\DeclareUnicodeCharacter{03B7}{\ensuremath{\eta}}
\DeclareUnicodeCharacter{03C0}{\ensuremath{\pi}}
\DeclareUnicodeCharacter{2192}{\ensuremath{\rightarrow}}
\DeclareUnicodeCharacter{2013}{--}
\DeclareUnicodeCharacter{2019}{'}
\usepackage{amsmath}
\usepackage{amssymb}
\usepackage{color}
\usepackage{authblk}
\usepackage[numbers,sort&compress]{natbib}
\usepackage[margin=1in]{geometry}
\usepackage[
    colorlinks=true,
    linkcolor=blue,
    citecolor=blue,
    urlcolor=blue
]{hyperref}

\usepackage{graphicx}
\DeclareGraphicsExtensions{.pdf,.png,.jpg}	

\title{Sgoldstino Phenomenology at SND@HL-LHC}

\author[1]{Kalashnikov D.}
\author[2]{Karkaryan E.}
\affil[1]{Institute for Nuclear Research of the Russian Academy of Sciences, Moscow 117312, Russia\\\texttt{kalashnikov@inr.ac.ru}}
\affil[2]{I.E. Tamm Department of Theoretical Physics, Lebedev Physical Institute, 53 Leninskiy Prospekt, Moscow, 119991, Russia\\\texttt{karkaryan@bk.ru}}
\date{}

\begin{document}

\maketitle

\begin{abstract}
We study the phenomenology of light scalar sgoldstinos, with masses from the dimuon threshold up to a few GeV, focusing on their production and detection prospects at the SND@HL-LHC experiment. We review the effective sgoldstino interactions with Standard Model particles, the dominant meson-decay production channels, and the relevant experimental constraints on the model parameter space. We also outline the recently proposed SND@HL-LHC detector configuration and discuss the kinematics and track separation of the muon pair produced in sgoldstino decays inside the detector volume. For two representative sets of supersymmetry-breaking parameters, we present sensitivity estimates and show that the muon--antimuon separation efficiency should be taken into account in searches for the dimuon signal.
\end{abstract}

\section{Introduction}

The Standard Model (SM) provides an accurate description of particle interactions at presently explored energies, but it leaves several conceptual and phenomenological questions open. Among them are the origin of the electroweak scale, the nature of dark matter, and the possible existence of additional weakly coupled sectors. Supersymmetry is one of the most developed extensions of the SM: it relates bosonic and fermionic degrees of freedom and can stabilize scalar masses against large radiative corrections~\cite{Wess:1974tw,Giudice:1998bp}. Since supersymmetry is not observed as an exact symmetry in nature, it must be broken. If supersymmetry breaking is associated with a hidden sector, it can give rise to new low-energy phenomenology. The corresponding Goldstone particle is called goldstino. In realistic low-scale scenarios, the goldstino belongs to a chiral supermultiplet whose scalar components are the scalar and pseudoscalar sgoldstinos~\cite{Brignole:1996fn,Gorbunov:2000th,Brignole:2003cm}. Assuming conserved $R$ parity, sgoldstinos can be produced singly in processes involving only SM particles~\cite{Perazzi:2000sgoldstino, gorbunov2001compheppackagelightgravitino}. Their masses and interactions depend on the scale and mediation mechanism of supersymmetry breaking, and they can be sufficiently light to be produced in meson decays or through couplings to SM gauge bosons and fermions~\cite{Brignole:2003cm,Gorbunov:2000th,Astapov:2015otc}. This makes light sgoldstinos a useful benchmark for feebly coupled new particles, especially in the mass range where intensity-frontier and forward-detector experiments are competitive with traditional high-$p_T$ searches.

The forward direction at the LHC provides a particularly suitable environment for such searches. A large flux of light and heavy mesons is produced at small angles with respect to the beam axis, and their rare decays can generate long-lived particles that travel hundreds of meters before decaying. SND@LHC and its proposed high-luminosity upgrade, SND@HL-LHC, are designed to exploit this forward region~\cite{Boyarsky:2022SND,Abbaneo:2895224,Abbaneo:2926288}. More broadly, the physics potential of SND@LHC and other far-forward LHC experiments has been explored for a wide range of new-physics scenarios, including long-lived particles and dark-sector models~\cite{Boyarsky:2022SND,Anchordoqui:2021ghd,Demidov:2022fas}. Here we adapt the sgoldstino phenomenology to the SND@HL-LHC geometry, focusing on scalar sgoldstino production in meson decays and on subsequent decays into muon--antimuon pairs inside the detector volume, with particular emphasis on the role of muon--antimuon separation in a magnetized calorimeter. In this respect, our study complements existing far-forward new-physics studies by examining more closely how the detector response to a dimuon final state affects the observable signal. The dimuon final state is especially attractive because it can yield a clean signature, provided that the two tracks can be separated and their charges identified.

We organize the paper as follows. First, we summarize the effective sgoldstino interactions and the relevant decay modes. We then discuss sgoldstino production mechanisms, the experimental constraints on these channels, and the SND@HL-LHC detector parameters relevant for our analysis. Finally, we evaluate the muon--antimuon separation factor and present the resulting sensitivity estimates for two representative choices of supersymmetry-breaking parameters.

\section{Sgoldstino interaction Lagrangian}

To study scalar sgoldstino phenomenology at energies below the electroweak scale, we consider a relevant part of an effective Lagrangian describing scalar sgoldstino interactions with Standard Model (SM) particles~\cite{Gorbunov:2000th,Perazzi:2000sgoldstino}. We note that, in contrast to the goldstino, which is $R$-odd, both scalar and pseudoscalar sgoldstinos are $R$-even~\cite{gorbunov2001compheppackagelightgravitino}. Consequently, interactions containing a single sgoldstino and only SM fields conserve $R$ parity and allow for single-sgoldstino production in SM processes. The complete effective Lagrangian also contains interactions of the pseudoscalar sgoldstino and terms involving two sgoldstinos \cite{Gorbunov:2000th}. The latter appear at higher order in the effective expansion and are suppressed by additional powers of $1/F$ in comparison with the single-sgoldstino interactions. Their contribution to sgoldstino production is therefore negligible in the parameter region studied in this work. We focus primarily on the scalar sgoldstino $S$, while the phenomenology of the pseudoscalar sgoldstino is briefly discussed at the end of the Results and Discussion section. The relevant effective Lagrangian then reads
\begin{equation} \label{eq:Lagr}
    {\cal L} = -\frac{M_{\gamma\gamma}}{2\sqrt{2}F}SF_{\mu\nu} F^{\mu\nu} 
    -\frac{M_3}{2\sqrt{2}F}SG^a_{\mu\nu} G^{\mu\nu\;a} - S\,\frac{v A_q y^q_{ij}}{\sqrt{2}\,F}\bar q_i q_j  
   -S\,\frac{v A_l y^l_{ij}}{\sqrt{2}\,F}\bar l_i l_j \, ,  
\end{equation}
where $F^{\mu \nu}$ denotes the photon field strength, $G^{\mu \nu}_a$, with $a=1,2,3,\dots, 8$, denotes the gluon field strength, $q_i$, with $i=1,\dots,6$, denote quarks, and $l_i$, with $i=1,2,3$, denote charged leptons. The supersymmetry-breaking parameter $F$ has dimension of mass squared, $M_{\gamma\gamma}\equiv M_1\cos^2\theta_W+M_2\sin^2\theta_W$, where $M_i$, $i=1,2,3$, are the gaugino masses corresponding to the three SM gauge groups $U(1)_Y$, $SU(2)_W$, and $SU(3)_c$, respectively, and $v=175$\,GeV is the Higgs vacuum expectation value. For the soft supersymmetry-breaking trilinear couplings entering~\eqref{eq:Lagr}, we use the approximation $v A_{q,l} y_{ij}^{q,l}$, where parameters $A_{q,l}$ define the common scale of the trilinear terms in the squark and slepton sectors. For flavor-conserving couplings, we assume the corresponding parameters to be proportional to the fermion masses, $v y_{ii}^{q,l}=m^{q,l}_i$. For flavor violation, we use the notation $v A_q y_{ij}^{q} = m_{q \, ij}^{LR \, 2}$.

\subsection{Sgoldstino decays} \label{S_decays}
We are primarily interested in sgoldstino decays into muon pairs, since this channel could provide a clean, background-free signal at SND. Sgoldstino decays into lepton pairs are governed by~\eqref{eq:Lagr}. The corresponding decay widths are
\begin{equation}\label{decay_leptons}
    \Gamma(S \rightarrow l^+ \, l^-) = \frac{m_{S} A_l^2 m_l^2}{16 \pi F^2} \times \Big(1-\frac{4m_l^2}{m_{S}^2}\Big)^{3/2}\,.
\end{equation}

\begin{figure}[ht]
    \centering
    \begin{minipage}[t]{0.49\textwidth}
        \centering
        \includegraphics[width=\linewidth]{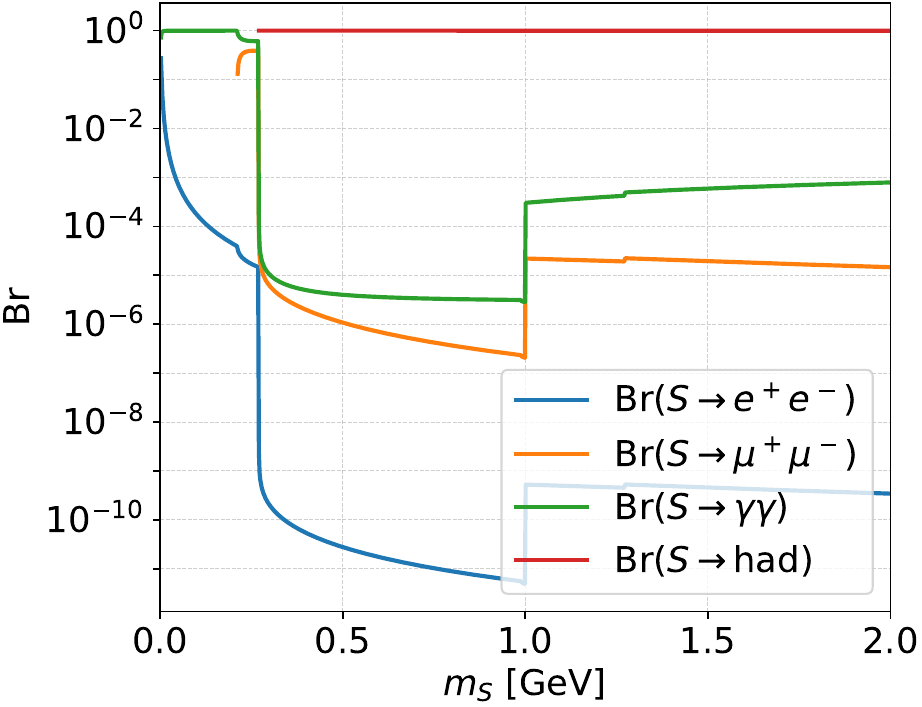}
        \par\small (a) set 1
    \end{minipage}\hfill
    \begin{minipage}[t]{0.49\textwidth}
        \centering
        \includegraphics[width=\linewidth]{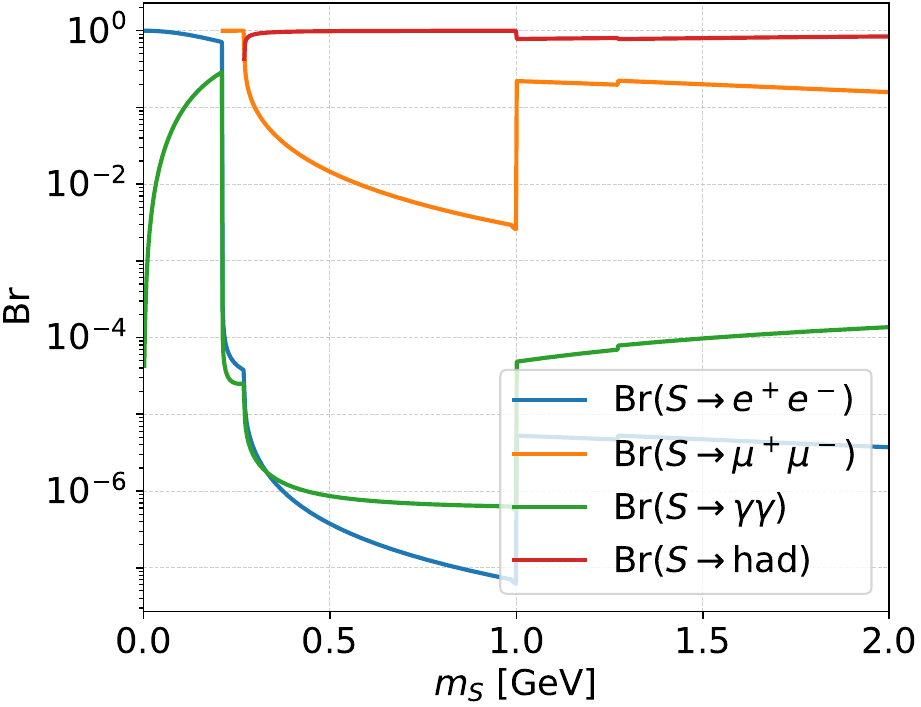}
        \par\small (b) set 2
    \end{minipage}
    \caption{Branching fractions of sgoldstino decays into lighter SM particles for the two model parameter sets \eqref{eq:sets} at $\sqrt{F}=500$ TeV.}
    \label{fig:branchings}
\end{figure}

The sgoldstino also decays into SM photons, with width
\begin{equation}\label{decays-to-photons} 
    \Gamma(S \rightarrow \gamma \, \gamma) = 
    R_\gamma \times 
    \frac{m_{S}^3 M_{\gamma \gamma}^2}{32 \pi F^2}\,.
\end{equation}
Here the factor $R_\gamma$ accounts for the scale dependence of the coupling in~\eqref{eq:Lagr}; see, e.g.,~\cite{Voloshin:1980zf,Gorbunov:2000th}. At one-loop level it has the form
\begin{equation}\label{rg_factor}
    R_\gamma = \left( \frac{b(m_S) \, \alpha(m_S)}{b(M_{\gamma \gamma}) \, \alpha(M_{\gamma \gamma})}\right)^2 \, ,
\end{equation}
with the electromagnetic coefficient $b(m_S) = \sum\limits_{m_f<m_S} N_c^f Q_f^2$. Here $N_c^f$ is the number of colors of the fermion $f$, and $Q_f$ is its electric charge. For $m_S$ up to $400$ MeV, only light charged fermions contribute to the factor $b$, giving $R_\gamma \approx 0.3$. In the relevant parameter region (see Eq.~\eqref{eq:sets} and Figs.~\ref{fig:opt_sens},~\ref{fig:cons_sens}), the SND@HL-LHC sensitivity is either restricted to $m_S<350$ MeV or the photon decay width is negligibly small, see Fig~\ref{fig:branchings} panel (b). We therefore treat $R_\gamma$ as a constant and set $R_\gamma = 0.3$.

Hadronic decays of light sgoldstinos can be described within chiral perturbation theory ($\chi$PT). In this regime, the dominant hadronic channels are scalar-sgoldstino decays into pions and kaons. Following~\cite{Gorbunov:2000th,Bezrukov_2010}, we introduce the gluon, quark, and Higgs-mixing contributions to the decay amplitude of $S \rightarrow \pi^0 \pi^0$,
\begin{eqnarray} \label{eq:S_M_decay_amp}
    A_g = - \frac{\alpha_s}{\beta(\alpha_s)} \times \frac{2\pi M_3m_S^2 }{\sqrt{2}F} = -\frac{M_3}{\sqrt{2}F} \times \frac{8\pi^2 m_S^2}{7\alpha_s(M_3)} \\
    A_q = -\frac{A_q}{\sqrt{2}F} \times m_\pi^2 \\
    A_h = - \frac{\theta}{\sqrt{2} v} \times \frac{2m_S^2 -11m_\pi^2}{9} \, ,
\end{eqnarray}
where $\alpha_s(M_3)$ is the strong coupling constant at the scale $M_3$. The mixing angle between the sgoldstino and the Higgs boson is given by~\cite{Astapov:2014mea}
\begin{equation}
\label{eq-mixing}
    \theta = -\frac{X}{Fm_h^2}\,,\;\;\;
    X = 2\mu^3 v \sin{2\beta} + \frac{1}{2}v^3(g_1^2M_1 + g_2^2M_2)\cos{2 \beta}\;,
\end{equation}
where $m_h=125$ GeV is the SM Higgs mass, $\mu$ is the SUSY Higgs-sector mass parameter, and $\tan \beta$ is the ratio of the Higgs fields.
For $\tan \beta = 6$, $\mu=1$ TeV, $g_1 = 0.349$, $g_2 = 0.654$, and $M_1 = M_2 = M_{\gamma \gamma}$, Eq.~\eqref{eq-mixing} yields the following expression for the mixing parameter:
\begin{equation}
    \frac{X}{1 \, \text{TeV}^4} \approx 0.11 - 0.0014  \times \frac{M_{\gamma\gamma}}{1 \, \text{TeV}},
\end{equation}
and for the mixing angle
\begin{equation}
    \theta = -\left( 7.3 - 0.09 \frac{M_{\gamma \gamma}}{1 \, \text{TeV}} \right) \frac{1 \, \text{TeV}^2}{F} \, .
\end{equation}
In the parameter region of interest (see Figs.~\ref{fig:opt_sens},~\ref{fig:cons_sens},~\ref{fig:sensitivity-fv} and Eq.~\eqref{eq:sets}), $|\theta| \sim 10^{-6}$--$10^{-3}$.

Using~\eqref{eq:S_M_decay_amp}, we obtain
\begin{equation}
    \Gamma (S \rightarrow \pi^0 \pi^0) = \frac{1}{16\pi  m_S} \times  |A_\text{tot}|^2 \times \sqrt{1-\frac{4m_\pi^2}{m_S^2}} \, ,    
\end{equation}
where $A_\text{tot} = A_g+A_q+A_h$. For the charged pions 
\begin{equation}
    \Gamma (S \rightarrow \pi^+ \pi^-) = 2\Gamma (S \rightarrow \pi^0 \pi^0) \, .
\end{equation}
Following the same method, we obtain the sgoldstino decay width into kaons
\begin{equation}
\begin{split}
    & \Gamma (S \rightarrow K^+ K^-) = \Gamma (S \rightarrow K^0 \bar{K^0}) = \frac{1}{16\pi  m_S} \times \sqrt{1-\frac{4m_K^2}{m_S^2}} \\
    & \times  \left( \frac{8 \pi^2 M_3 m_S^2}{7F \, \alpha_s(M_3)} + \frac{A_q m_K^2}{F} + \frac{\theta}{v} \left( \frac{2m_S^2 -11m_K^2}{9} \right) \right)^2 \, ,
\end{split}    
\end{equation}

For heavy sgoldstinos, the hadronic mode can be described by the decay into a gluon pair:
\begin{equation}
\label{rate-to-gluons}
    \Gamma(S \rightarrow g \, g) = R_g \times \frac{{m_S}^3 M_3^2}{4 \pi F^2} \, ,
\end{equation}
Here $R_g$ is a rescaling factor analogous to that in~\eqref{rg_factor}:
 \begin{equation}\label{rg_factor_gluons}
    R_g = \left( \frac{b_0(m_S) \, \alpha_s(m_S)}{b_0(M_3) \, \alpha_s(M_3)}\right)^2 \, ,
\end{equation}
where $b_0(m_S) = 11 - 2n_f/3$, and $n_f(m_S)$ is the number of quark flavors with masses $m_f<m_S$.

We use $m_S = 1.0$ GeV as the threshold for the transition between the meson-width and gluon-width descriptions. In this mass region, chiral perturbation theory has sizable uncertainties, and additional resonances can contribute. Since most of the SND@HL-LHC reach obtained below (see Figs.~\ref{fig:opt_sens},~\ref{fig:cons_sens}) covers models with $m_S<400$ MeV, we do not investigate these uncertainties further. Sensitivity regions also appear for heavier sgoldstinos with $m_S \sim 1$ GeV; in this region, as stated above, we include only the gluonic description of the hadronic decay modes.

In a natural supersymmetric model, all soft parameters are expected to be of the same order, e.g. $M_3 \sim M_{\gamma \gamma} \sim A_q \sim A_l < \sqrt{F}$. However, if $M_3 \sim M_{\gamma \gamma} \sim A_q \sim A_l$, decays into muons are subdominant throughout the sgoldstino mass range considered here. We therefore consider two phenomenologically motivated parameter sets,
\begin{equation} \label{eq:sets}
\begin{split}
    & \text{set 1: } M_3=M_{\gamma \gamma}=A_q=A_l= 0.3 \, \sqrt{F} \\
    & \text{set 2: } M_3=3 \text{ TeV}, \, M_{\gamma \gamma}=1 \text{ TeV}, \, A_q=A_l= 0.5 \, \sqrt{F} \, 
\end{split}
\end{equation}
with $\mu=1$ TeV and $\tan \beta = 6$ for both sets. The corresponding branching fractions are shown in Fig.~\ref{fig:branchings}. The values in the second set are chosen to keep the sgoldstino decay width small enough for sgoldstinos to reach the detector while maintaining large muon branching fractions and production rates. The feature at $m_S=1$ GeV is an artifact of our modeling choice to switch by hand at this mass from the $\chi$PT treatment to the gluonic treatment of the hadronic decay modes.

\subsection{Sgoldstino production mechanisms} \label{S_prod}

Mesons produced in proton collisions can decay into a sgoldstino if this decay is kinematically allowed. Many relevant production modes were investigated in Refs.~\cite{Gorbunov:2000th,Astapov:2015otc}.

\begin{figure}[!ht]
\centering
\includegraphics[width=0.95\linewidth]{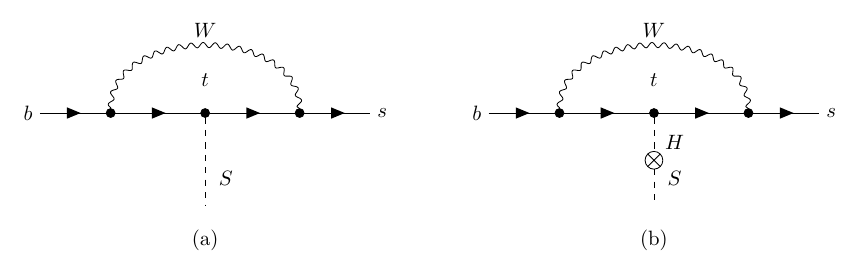}
\caption{Feynman diagrams for meson decays into a sgoldstino via flavor-conserving sgoldstino couplings\,\cite{Gorbunov:2000th}: (a) sgoldstino production through a Yukawa-type coupling to the top quark; (b) production through sgoldstino--Higgs mixing.}
\label{fig:feynman-higgs}
\end{figure}

\paragraph{Pseudoscalar meson decays.}
Decays into a sgoldstino and mesons with different quark flavor arise at one-loop level, with a virtual $W$ boson mediating the flavor change. Sgoldstinos are produced either through direct couplings to top quarks running in the loop or through mixing with the Higgs boson, which couples to the same quarks; see Refs.~\cite{Astapov:2015otc,Bezrukov_2010}. Examples of the corresponding Feynman diagrams for the $b \to s$ transition are shown in Fig.~\ref{fig:feynman-higgs}.
For beauty-meson decays into a strange meson and a sgoldstino, the branching ratio reads~\cite{Astapov:2015otc}
\begin{equation}\label{eq_BMesonDecay}
    \text{Br}(B \rightarrow X_s S)=0.3 \times \left( \frac{m_t}{m_W} \right)^4 \times \left( 1 - \frac{m_S^2}{m_b^2} \right)^2 \times \left( \frac{A_qv}{F} + \theta \right)^2
\end{equation}

Similarly, charmed and strange mesons can also decay into a sgoldstino. Following~\cite{Bezrukov_2010,Shifman1990}, we arrive to the kaon decay widths
\begin{equation} 
    \Gamma(K \rightarrow \pi S) = \left( \frac{3 G_F \, m_K^2}{32 \pi^2 \, v} V_{td}^*V_{ts}m_t^2 \right)^2 \times \frac{|\Bar{p}_S|}{8 \pi m_K^2} \times \left( \frac{A_qv}{F} + \theta \right)^2 \, ,
\end{equation}
where $K$ corresponds to $K^\pm$ and $K_L$. Here $G_F$ is the Fermi constant, $V_{ij}$ is a CKM-matrix element, and the sgoldstino three-momentum is
\[
|\Bar{p}_S|=\frac{1}{2m_K}\sqrt{\left( \left( m_K-m_S\right)^2-m_\pi^2\right)\left( \left( m_K+m_S\right)^2-m_\pi^2\right)}\,.
\]
The $K_S \to \pi^0 S$ amplitude is proportional to the imaginary part of the relevant CKM combination, and consequently, the $K_S$ contribution is strongly suppressed~\cite{Bezrukov_2010,Shifman1990}. Its branching fraction is further decreased because the total width of the short-lived $K_S$ is more than two orders of magnitude larger than those of $K^\pm$ and $K_L$~\cite{Zyla:2020zbs}. The contributions from $K^\pm$ and $K_L$ decays are instead limited by the probability that these long-lived parent mesons decay before reaching the relevant beamline material, which is of order $10^{-3}$ for kaons with momentum $p_K=10$ GeV. The parent kaons capable of producing sgoldstinos that reach the detector must have even larger momenta, and their decay probability is therefore further suppressed. Numerically, the total kaon contribution remains below three expected signal events throughout the parameter space considered.

Diagrams similar to those in Fig.~\ref{fig:feynman-higgs} also give rise to the decay $\eta\rightarrow \pi^0 S$. The order-of-magnitude branching fraction is given by~\cite{Shifman1990} 
\begin{equation}
    \text{Br}(\eta \rightarrow \pi^0 S) \sim 10^{-6} \times \frac{2|\Bar{p}_S|}{m_\eta} \times \left( \frac{A_qv}{F} + \theta \right)^2.
\end{equation}

The charmed-meson branching fractions are much stronger suppressed because of the relevant combinations of quark-mixing parameters and the smallness of the $b$-quark Yukawa coupling entering the loop amplitude. The largest contribution for charmed mesons comes from the tree-level process in which the sgoldstino is emitted from the initial quark, with branching fraction~\cite{Bezrukov_2010}
\begin{equation}
    \text{Br}(D \rightarrow e \nu \, S) \sim 10^{-8} \times \left( \frac{A_qv}{F} + \theta \right)^2 \, ,
\end{equation}
In the relevant region of parameter space, $\theta \lesssim 10^{-3}$, the resulting branching fraction is $\text{Br}(D \rightarrow e \nu \, S) \sim 10^{-14}$. The actual fraction of sgoldstinos from $D$-meson decays is even smaller because of kinematic factors. Given the expected number of produced $D$ mesons, this contribution is negligible. The same is true for sgoldstinos produced in $\eta$-meson decays, see Fig.~\ref{eq_BMesonDecay}. In both cases, the contribution to the number of signal events for the expected proton-collision statistics at the HL-LHC is much smaller than 1.

We now consider sgoldstino production channels initiated by flavor-violating sgoldstino interactions with quarks. The corresponding coupling constants are proportional to the off-diagonal entries of the left-right squark mass-squared matrix
\[
A_q v y^q_{ij}\equiv \tilde{m}_{ij}^{LR \; 2}\, .
\]
The sgoldstino is neutral, and hence the mixing involves either two up-type quarks or two down-type quarks. The decay rate of a pseudoscalar meson $M$ into a lighter meson $X$ and a scalar sgoldstino is~\cite{Demidov:2022fas}
\begin{equation}\label{eq:meson_decay}
    \Gamma(M \rightarrow XS) = F_{M \rightarrow X}^2 \left( \frac{m_M^2-m_X^2}{m_{q_i}-m_{q_j}} \right)^2 \frac{\sqrt{\lambda(m_M^2, m_X^2, m_S^2})}{32 \pi^2 m_M^3} \frac{\tilde{m}_{q \, ij}^{LR \, 4}}{F^2}\,.
\end{equation}
Here $\lambda(x,y,z)=x^2+y^2+z^2-2xy-2xz-2yz$ is the Källén function. The quantities $m_{q_i}$ are the quark masses, and $F_{M \rightarrow X}$ denotes a dimensionless form factor. The latter can be calculated for various meson transitions using the form factors of Ref.~\cite{Palmer_2014}:
\begin{equation}
    F_{M \rightarrow X}(0) = \frac{F_{D \rightarrow X}(0) \times m_D^{3/2}}{m_M^{3/2}}\,,
\end{equation}
\begin{equation}\label{eq67}
    F_{M \rightarrow X}(m_S^2) = \frac{F_{M \rightarrow X}(0)}{\left( 1 - \frac{m_S^2}{m_{M^*}^2} \right)\left( 1 - \frac{\tilde{\alpha} m_S^2}{m_{M^*}^2} \right)}\,,
\end{equation}
where $m_D=1.87$ GeV is the $D$ meson mass. The expression~\eqref{eq67} includes a pole at the mass of the first excited state $m_{M^*}$ and another one at $m_{M^*}/\sqrt{\tilde{\alpha}}$, where the dimensionless factor $\tilde{\alpha}=0.4$ accounts for contributions of higher resonances.

Using Eq.~\eqref{eq:meson_decay}, one can obtain the partial decay widths for $K \to \pi S$ and $K_S \to \pi S$ through $\tilde{m}_{d \; 12}^{LR}$, $B\to K S$ through $\tilde{m}_{d \; 23}^{LR}$, $B\to \pi S$ through $\tilde{m}_{d \; 13}^{LR}$, and $D\to \pi S$ and $D_s\to K S$ through $\tilde{m}_{u \; 12}^{LR}$. In our estimates, we take all $\tilde{m}_{q \, ij}^{LR}$ to be equal and set $\tilde{m}^{LR} = 30$ GeV. In the relevant region of parameter space, this satisfies the constraints on $\delta^q_{ij}=\frac{\tilde{m}^{q\,LR\,2}_{ij}}{m_{\tilde q}^2}<10^{-3}$~\cite{Crivellin:2008mq}, where the squark mass satisfies $m_{\tilde q} \approx M_3$, and provides sufficiently large branching fractions.

\begin{figure}[!ht]
    \centering
    \includegraphics[width=0.7\textwidth]{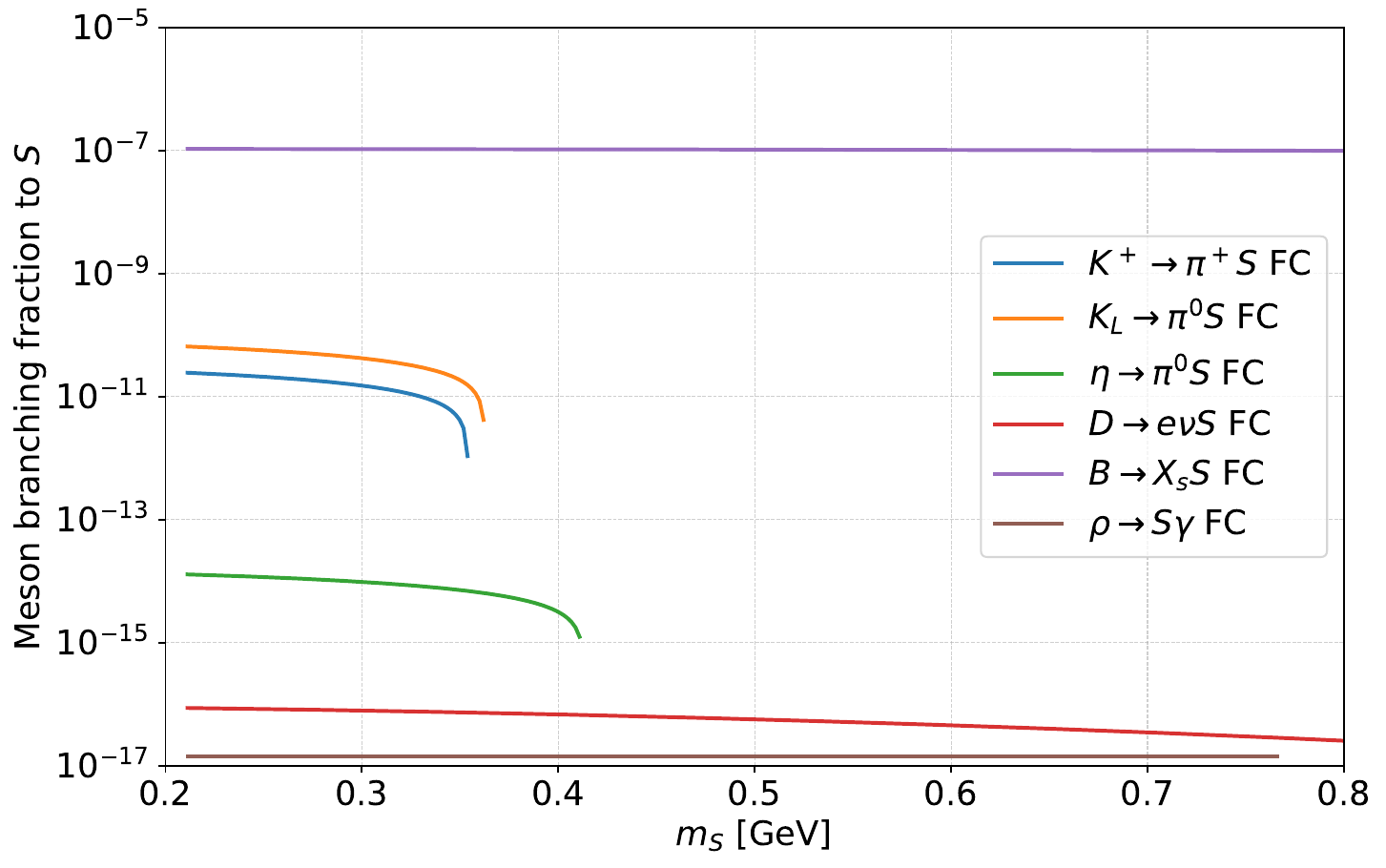}
    \caption{Comparison of parent-meson branching fractions into sgoldstinos for set 1 in~\eqref{eq:sets} and $\sqrt{F} = 500$ TeV. The FC tag indicates that only flavor-conserving sgoldstino couplings are included $(\tilde{m}^{LR}=0)$.}
    \label{fig:S_prod}
\end{figure}

\paragraph{Vector meson decays.}
Flavor-neutral vector mesons, such as $\rho^0$, $\phi$, $\omega$, $J/\psi$, and $\Upsilon$, can decay into a sgoldstino and a photon. Following Ref.~\cite{Gorbunov:2000th}, the corresponding branching fractions can be evaluated from the ratio of these rates to the rates into lepton pairs:
\begin{equation}
    \frac{\text{Br}(V \rightarrow S \gamma)}{\text{Br}(V \rightarrow \gamma \rightarrow e^+ e^-)} = \frac{M_V^2(A_q - M_{\gamma \gamma} \, R_\gamma)^2}{16 \pi \alpha F^2}\,,
\end{equation}
where $\alpha$ is the fine-structure constant. Over the relevant region of parameter space, $\text{Br}(V \rightarrow S \gamma) < 10^{-14}$. Together with kinematic factors, this results in a negligible contribution to the number of signal events. The relevant branching fractions are shown in Fig.~\ref{fig:S_prod} for one representative value of $\sqrt{F}$.

\paragraph{Direct production}
Another possible production mechanism is direct sgoldstino production through gluon fusion. This channel is controlled by the gluino-mass parameter $M_3$ and can therefore be especially relevant for sgoldstino masses between the dimuon and two-pion thresholds, $2m_\mu < m_S < 2m_\pi$, where hadronic sgoldstino decays are still kinematically closed. However, in this low-mass region the use of parton distribution functions is affected by sizable uncertainties, which can substantially change the predicted production rate. In addition, sgoldstinos produced directly in proton--proton collisions are expected to have larger momenta than those produced in the meson decays considered here. As discussed below, such a harder spectrum makes the two muon tracks more collimated and therefore more difficult to separate experimentally. For these reasons, direct sgoldstino production requires a dedicated analysis and is left for future work, once the final detector configuration and the appropriate treatment of the relevant parton distributions are clarified.

\paragraph{Direct limits on meson decays into sgoldstino.}

We conclude the discussion of sgoldstino production modes by noting that many of the channels above are constrained by dedicated studies of meson decays, which typically include searches for new light particles or missing-energy signatures. A summary of the upper limits used in this paper is presented in Table~\ref{tab:Br_constraints}.
\begin{table}[htb!]
    \begin{center}
    \begin{tabular}{| p{0.30\textwidth} | p{0.28\textwidth} | p{0.31\textwidth} |}
    \hline
    B-mesons & K-mesons & D-mesons \\
      \hline
    $\text{Br}(B \rightarrow X_s S)< 10^{-5}$ & $\text{Br}(K^\pm \rightarrow \pi^\pm S)<10^{-10}$ & $\text{Br}(D^\pm \rightarrow \pi^\pm S)<1.4\cdot 10^{-8}$\\
    $\text{Br}(B^0 \rightarrow \pi^0 S)<3.8 \cdot 10^{-8}$ & $\text{Br}(K_S\rightarrow \pi^0 S)<10^{-10}$ & $\text{Br}(D^0 \rightarrow \pi^0 S)<2.0\cdot 10^{-4}$\\ 
    $\text{Br}(B^\pm \rightarrow \pi^\pm S)<1.4 \cdot 10^{-5}$ & & $\text{Br}(D_s \rightarrow K S)<2.6\cdot 10^{-4}$\\
    \hline
    \end{tabular}
    \end{center}
\caption{Upper limits on the most relevant for our study meson branching fractions into sgoldstinos~\cite{Zyla:2020zbs,Gorbunov:2000th,PhysRevD.79.092004,Wei_2009}.}
\label{tab:Br_constraints}
\end{table}
In our analysis, we always consider regions of parameter space consistent with the bounds in Table~\ref{tab:Br_constraints}. When determining the SND reach, we include all kinematically open decay channels included in our effective model.

\section{SND@HL-LHC experiment}
\begin{figure}[!ht]  
\centerline{\includegraphics[width=0.5\linewidth]{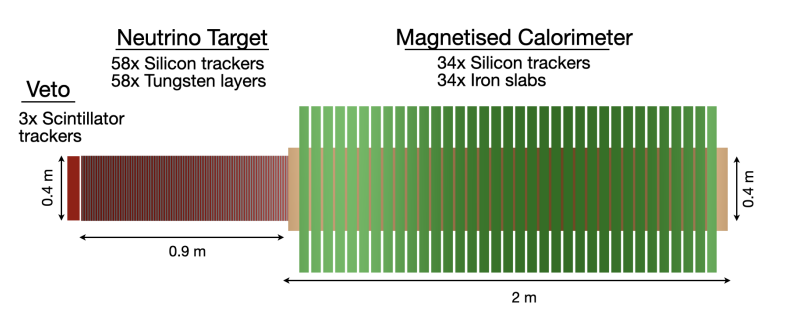} 
\includegraphics[width=0.3\linewidth]{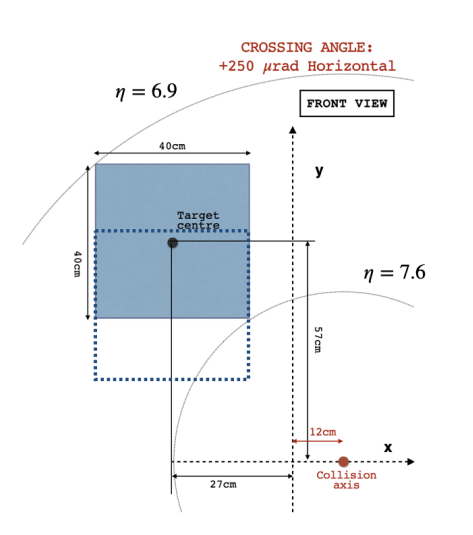} }
\caption{Outline of the proposed detector for SND@HL-LHC.}
\label{fig:advSND}
\end{figure}

SND@HL-LHC is a planned upgrade of SND@LHC, which has already demonstrated successful operation during Run~3. The authors of Ref.~\cite{Abbaneo:2926288} propose the installation of a magnetized hadronic calorimeter. In this section, we illustrate the importance of this modification for dimuon signature in searches for new physics. An outline of the proposed detector is shown in Fig.~\ref{fig:advSND}. SND@HL-LHC will be located at the site of the current SND@LHC detector, 480 m downstream of the ATLAS interaction point (IP). The detector will consist of an initial veto layer, which rejects most SM particles originating from the IP, followed by a 0.9 m long neutrino target and a magnetized hadronic calorimeter (HCAL) with magnetic-field strength $B=1.75$ T along the $y$ direction with respect to the detector axis $z$. The HCAL consists of 34 iron slabs, each 5 cm thick and followed by a 0.05 cm thick sensitive silicon layer. The main purpose of the calorimeter is to enable muon/antimuon charge identification, which in turn is crucial for neutrino/antineutrino identification. We use the detector position and size proposed in Ref.~\cite{Abbaneo:2926288}: the detector center is located at $\tilde{x}_0=27$ cm and $\tilde{y}_0=57$ cm with respect to the beam axis $\tilde{z}$, and the detector has a square cross section with side length $a=40$ cm. Here we study the range of model parameters for which the magnetized calorimeter helps to distinguish a single-muon event from a muon--antimuon pair, making decays of new-physics particles into muon--antimuon pairs a clean experimental signature.

In our numerical modeling, we use the following detector parameters:
\begin{table}[!htb]
\begin{center}
\begin{tabular}{| c | c | c | c | c | c |}
    \hline
    $ D $ & $L_{det}$ & $B$ & a & $\tilde{x}_0$ & $\tilde{y}_0$ \\
    \hline
    480 m & 2.9 m & 1.75 T & 40 cm & 27 cm & 57 cm  \\
    \hline
\end{tabular}
\end{center}
\caption{Summary of the relevant detector parameters~\cite{Abbaneo:2926288}.}
\label{tab:detector} 
\end{table}

\subsection{Muon--antimuon separation} \label{Muon_sep}
To treat sgoldstino decays into muon--antimuon pairs as a background-free signature, we must distinguish muon--antimuon pairs from single energetic muon events that might originate from neutrino interactions in the target. For this, muons and antimuons must be separated by a measurable distance, so that two distinct tracks can be observed inside the detector.

\begin{figure}[!ht]  
    \centering
    \begin{minipage}[t]{0.49\textwidth}
        \centering
        \includegraphics[width=\linewidth]{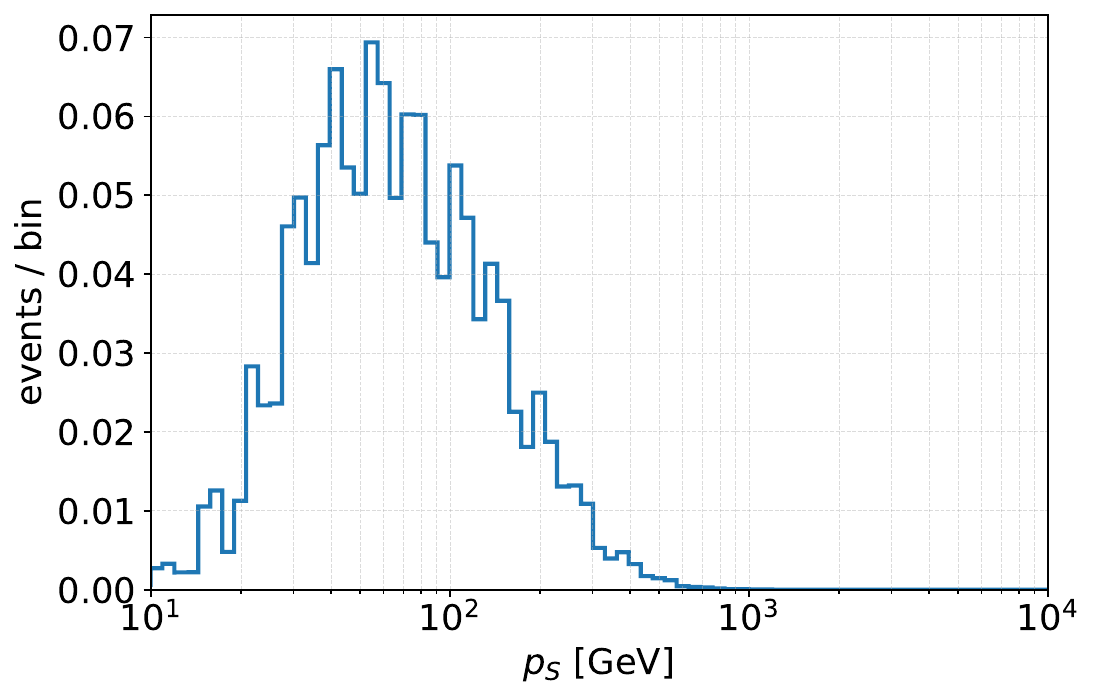}
        \par\small (a) $K$ channel. $m_S=0.25$ GeV, $\sqrt{F}=500$ TeV.
    \end{minipage}\hfill
    \begin{minipage}[t]{0.49\textwidth}
        \centering
        \includegraphics[width=\linewidth]{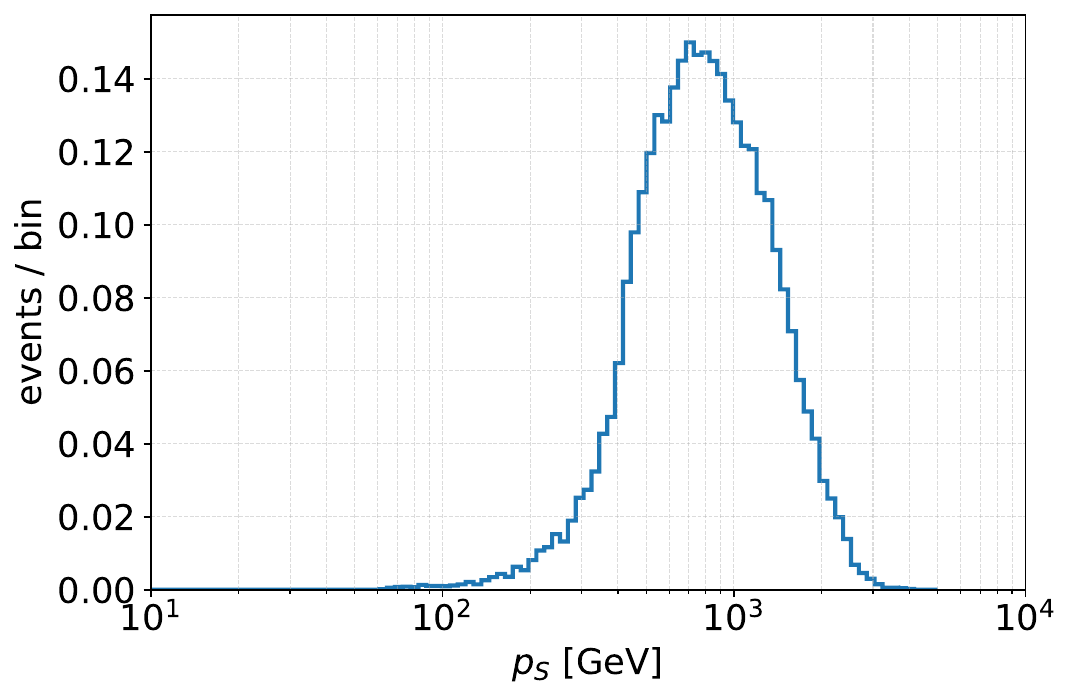}
        \par\small (b) $B$ channel. $m_S=0.25$ GeV, $\sqrt{F}=500$ TeV.
    \end{minipage}
\caption{Sgoldstino momentum distributions for decays inside the detector volume. Set~2 parameters from Eq.~\eqref{eq:sets} are used.}
\label{fig:Sspectra}
\end{figure}

Due to the the high momentum, the muon kinematics is governed mainly by the kinematics of the parent sgoldstino. Examples of momentum spectra for sgoldstinos decaying inside the decay volume are presented in Fig.~\ref{fig:Sspectra}.
The larger average sgoldstino momentum for $B$ mesons follows from the larger momentum of the parent meson. For forward meson production, the characteristic transverse-momentum scale is $p_T\sim m_B$ for $B$ mesons~\cite{Kling:2021fwx}; therefore, in order to have a sufficiently small angle to point toward the SND detector, the meson must have larger $p_z$, resulting in larger overall momentum.

After the sgoldstino decay, the muons have velocity $\beta$ and form an angle $\phi$ with the magnetic field and an angle $\alpha$ with the detector axis $z$. The deviation of the muon trajectory inside the magnetic field is then described as follows:
\begin{eqnarray}
    & x = \frac{\beta \sin{\phi}}{\omega} \left[ \cos{\alpha} - \cos \left( \alpha+ \omega t \right) \right] \\
    & y = \beta \cos\phi\,t \, \\
    & z =  \frac{\beta \sin{\phi}}{\omega} \left[ \sin \left( \alpha+ \omega t \right) -  \sin{\alpha} \right] \, ,
\end{eqnarray}
where $E$ is the muon energy and $\omega = \frac{\pm eB}{E}$, with "$+$" for muons and "$-$" for antimuons. Instead of evaluating the muon and antimuon coordinates analytically, we numerically propagate muons from sgoldstino decays through the detector and evaluate the distance between them at different values of $z$. The distance between the two muons at $z$ is then
\begin{equation}
    \Delta r^2 (z) = \Delta x^2 (z) + \Delta y^2 (z).
\end{equation}

\begin{figure}[ht]
    \centering
    \includegraphics[width=0.75\linewidth]{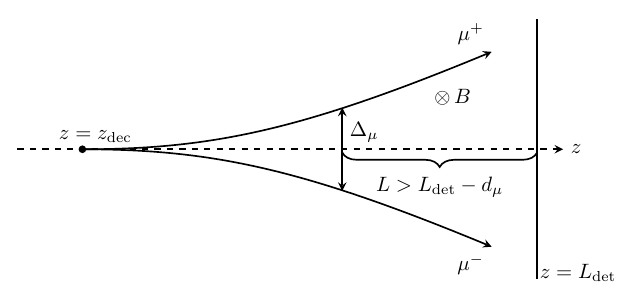}
    \caption{Schematic illustration of the muon--antimuon separation criterion used in this work. The event is accepted if the transverse distance between the two tracks exceeds $\Delta_\mu$ over a path length larger than $d_\mu$. The vertical line at $z=L_{\rm det}$ illustrates the downstream end of the detector. The point at $z=z_{\rm dec}$ shows the sgoldstino decay point. The position $z=0$ corresponds to the target surface facing the IP.}
    \label{fig:muon-separation-scheme}
\end{figure}

A sgoldstino decay is accepted if the distance between the two tracks exceeds $\Delta_\mu$ over at least a path length $d_\mu$ along the $z$ axis inside the detector. This separation criterion is illustrated schematically in Fig.~\ref{fig:muon-separation-scheme}. Upon integrating the muon-pair distribution over angles, we evaluate the muon-separation factor $A_\mu(m_S, p_S, z_{\rm{dec}})$ for a sgoldstino with mass $m_S$, momentum $p_S$ and decay point at $z_{\rm{dec}}$. This factor is the probability that a sgoldstino decay into a muon--antimuon pair produces two tracks that can be resolved by the HCAL.

\begin{figure}[ht]
    \centering
    \begin{minipage}[t]{0.49\textwidth}
        \centering
        \includegraphics[width=\linewidth]{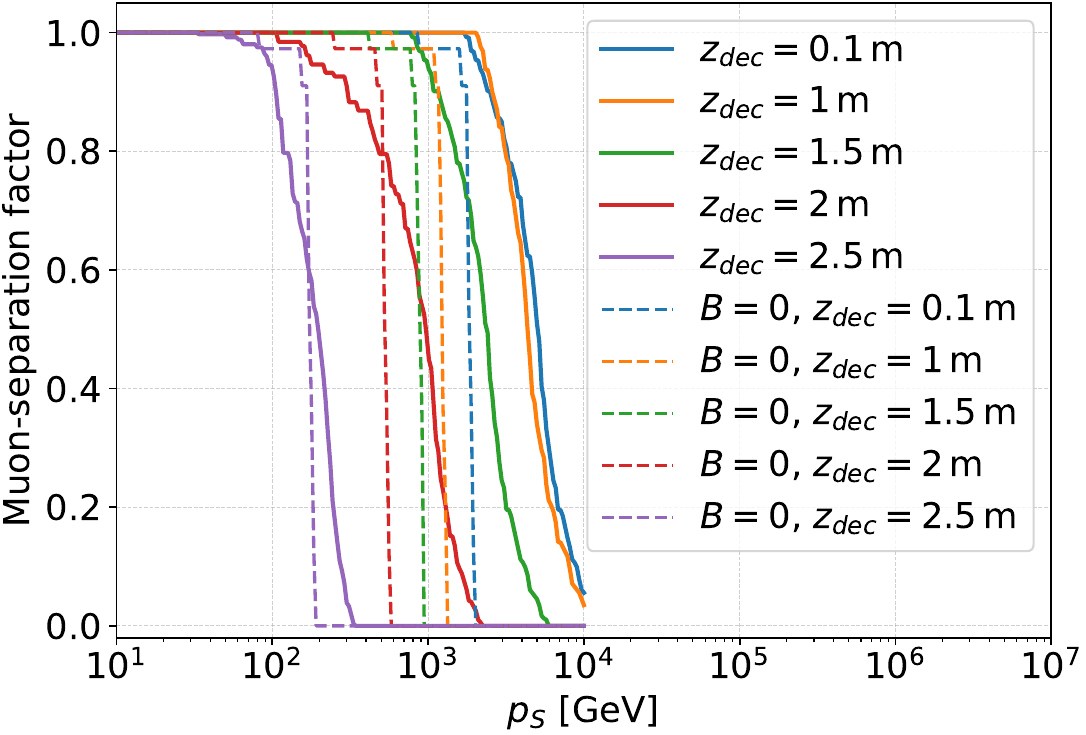}
        \par\small (a) Optimistic $\Delta r>1$ mm
    \end{minipage}\hfill
    \begin{minipage}[t]{0.49\textwidth}
        \centering
        \includegraphics[width=\linewidth]{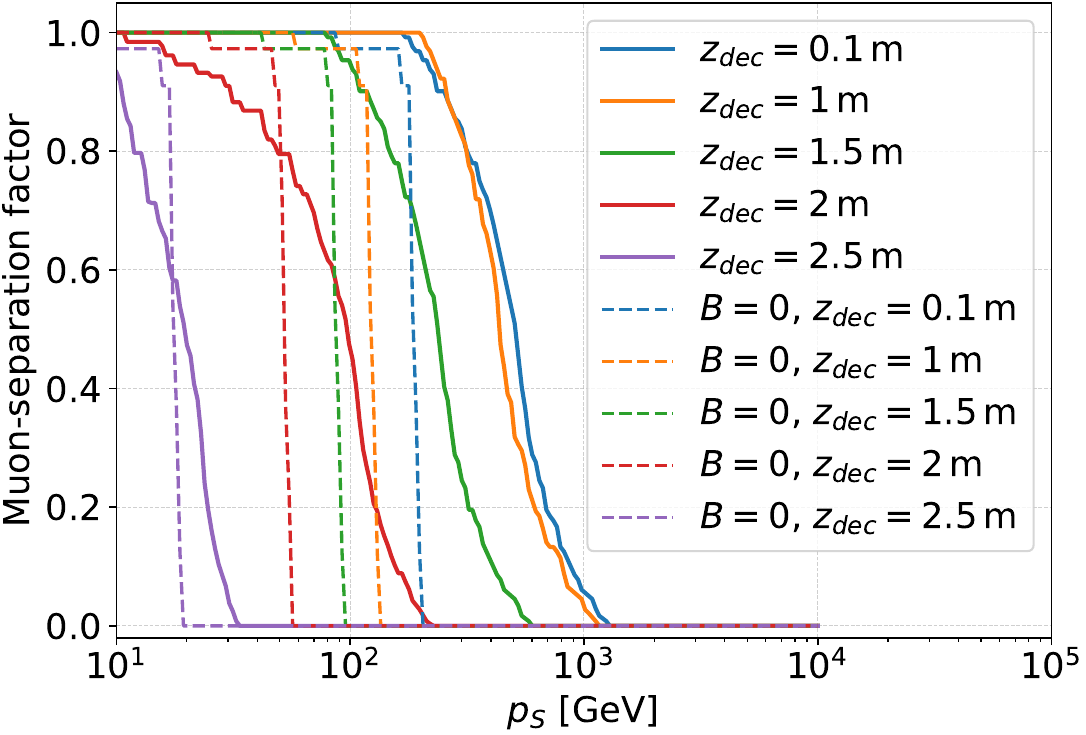}
        \par\small (b) Conservative $\Delta r>1$ cm
    \end{minipage}
    \caption{Fraction of muon--antimuon pairs satisfying the cut for different sgoldstino decay points, assuming that the initial sgoldstino direction is aligned with the $z$ axis. $m_S=400$ MeV}
    \label{fig:muon_factor}
\end{figure}

For our estimates, we require the minimum distance between the muon and antimuon in a sensitive HCAL layer to exceed $\Delta_\mu$ in order for the two tracks to be resolved. To allow track reconstruction, the lepton pair should cross several sensitive detector layers. Motivated by the HCAL segmentation described in Ref.~\cite{Abbaneo:2926288}, we therefore take $d_\mu = 15$ cm, corresponding to three 5-cm-thick layers, and require $\Delta r > \Delta_\mu$ over this path length. We then evaluate the fraction of muon--antimuon pairs that satisfy this requirement for sgoldstinos with momenta distributed as in Fig.~\ref{fig:Sspectra}. Since no dedicated study of muon--antimuon separation in the SND HCAL is available, we adopt the spatial-resolution benchmarks proposed in Ref.~\cite{Abbaneo:2926288}, where several values of $\Delta_\mu$ for single-muon reconstruction were considered. We take $\Delta_\mu = 1$ mm as an optimistic criterion and $\Delta_\mu = 1$ cm as a conservative one. The results are presented in Fig.~\ref{fig:muon_factor} for several sgoldstino decay coordinates inside the detector $z_{dec}$.

\begin{figure}[ht]
    \centering
    \includegraphics[width=0.55\linewidth]{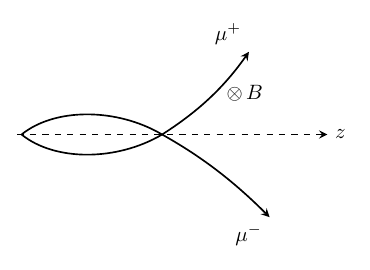}
    \caption{Schematic illustration of the muon--antimuon trajectory for opposite bending and the corresponding initial muon momentum directions.}
    \label{fig:muon-bending-scheme}
\end{figure}

As shown in Fig.~\ref{fig:muon_factor} the muon-antimuon separation factor behaves like a step function on the sgoldstino momentum. For sgoldstinos with momentum exceeding 1 TeV it becomes very difficult to separate muons and antimuons with the conservative threshold. This implies that large portion of sgoldstinos produced in $B$-meson decays do not satisfy the dilepton-separation requirement.
Two points are worth discussing here. First, the overlap of the $z_{dec}=1$ m and $z_{dec}=0.1$ m curves is explained by the non-magnetized neutrino-target volume before the magnetized calorimeter. For high-momentum sgoldstinos, the muon--antimuon pair is highly collimated. Therefore, in the absence of a magnetic field, the distance between them changes very little over the relevant timescales. Since the target volume is not magnetized, there is essentially no difference between highly collimated muons produced at $z=0$ and at $z=0.9$ m.
Second, the muon-separation factor can be larger for a non-magnetized calorimeter in the case of low-momentum muons. Since the magnetic field is directed along the $y$ axis, it bends muon trajectories toward $+x$ and antimuon trajectories toward $-x$. If the initial muon and antimuon momentum directions are opposite, namely $p^{\mu^-}_x<0$ and $p^{\mu^+}_x>0$, then the magnetic field can actually cause the muon and antimuon tracks to cross as shown schematically in Fig.~\ref{fig:muon-bending-scheme}.

\section{Numerical scheme}

In this section, we describe the calculation of the expected number of signal events. As discussed in Sec.~\ref{S_prod}, we consider sgoldstinos produced in meson decays. First, we use the EPOS-LHC/CRMC~\cite{Pierog_2015,Ulrich:2021crmc} and PYTHIA~8~\cite{Sjostrand:2014zea} packages to generate samples of mesons produced at the HL-LHC. For light mesons, we use EPOS-LHC/CRMC, whereas for heavy $B$ mesons we use PYTHIA. For each meson sample, we construct a binned distribution of the meson direction and momentum, $dn_M$. Then, for $m_S \in [2m_\mu; m_b]$, we evaluate the sgoldstino kinematic distribution normalized to the number of $pp$ collisions after the parent-meson decay in terms of $\beta \gamma$, retaining only sgoldstinos traveling inside the detector, $dn_S^{(M)}$. Thus, only sgoldstinos with momentum $(p_{\tilde{x}},p_{\tilde{y}},p_{\tilde{z}})$ satisfying Eq.~\eqref{eq:kin_cuts} are retained:
\begin{equation} \label{eq:kin_cuts}
\begin{split}
        & p_{\tilde{z}} > 0 \\
        & |\tilde{x}_S - \tilde{x}_0| \leq a/2 \\
        & |\tilde{y}_S - \tilde{y}_0| \leq a/2 \, ,
\end{split}
\end{equation}
where, for the sgoldstino,
\begin{equation}
\begin{split}
        & \tilde{x}_S = \frac{p_{\tilde{x}}}{p_{\tilde{z}}}D \\
        & \tilde{y}_S = \frac{p_{\tilde{y}}}{p_{\tilde{z}}}D \, ,
\end{split}
\end{equation}
where $(0,0,\tilde{z})$ denotes the beam axis, $(\tilde{x}_0,\tilde{y}_0, 0)$ is the position of the center of the detector's front surface, $a$ is the detector side length, and $D$ is the distance between the IP and the detector; see Table~\ref{tab:detector}.

After this kinematic precomputation, we scan over a grid in $(m_S,F)$ for the two sets of model parameters in Eq.~\eqref{eq:sets}. For each point on the $(m_S,F)$ grid and for each set of model parameters, we evaluate the branching fractions of meson decays into a sgoldstino, $\text{Br}(M \rightarrow S)$, the total width of the sgoldstino, $\Gamma_S$, and the branching fraction of $S$ into a muon pair, $\text{Br}(S \rightarrow \mu^+ \mu^-)$.

Using the sgoldstino distribution $dn_S^{(M)}/d(\beta\gamma)$, we evaluate the decay length $\lambda_S = \frac{\beta \gamma}{\Gamma_S}$. Here we assume that all sgoldstinos satisfying~\eqref{eq:kin_cuts} travel along the detector axis. Then the probability for the sgoldstino to decay inside the detector volume at distance $z$ is
\begin{equation}
    dP(\lambda,z) = \frac{dz}{\lambda_S} \exp\left[-\frac{D+z}{\lambda_S}\right] .
\end{equation}

Applying the muon-separation factor $A_\mu$ from Sec.~\ref{Muon_sep}, we arrive at the following expression, summed over all parent mesons $M$:
\begin{equation}
\begin{split}
    N_S(m_S,F) = N_{pp} \sum_M \int dz \, \frac{dn_S^{(M)}}{d(\beta \gamma)} & \times \text{Br}(M \rightarrow S)\,\text{Br}(S \rightarrow \mu^+ \mu^-) \\
    & \times dP(\lambda,z) A_\mu(m_S,p_S,z) \, ,
\end{split}
\end{equation}
where $N_{pp}=2.26 \cdot 10^{17}$ is the expected number of $pp$ collisions at the HL-LHC. We obtain this number using the inelastic $pp$ cross section $\sigma_{\mathrm{inel}}=75.4$~mb~\cite{LHCb:2018uhs} and the HL-LHC integrated luminosity $\mathcal{L}_{\text{int}}=3000\;\mathrm{fb}^{-1}$. 

After applying the geometrical acceptance, the decay-inside-the-detector requirement, and the muon-separation factor, we treat the selected two-track dimuon signature as background free. Background estimates for visible long-lived-particle decays at far-forward LHC experiments indicate that the residual background after veto and displaced-vertex requirements is very small; consequently, such signatures can be treated as effectively background-free in sensitivity projections for SND@LHC~\cite{Feng_2018,Ariga_2019,Boyarsky:2022SND,Demidov:2022fas}. Under this assumption, we use the 95\% CL sensitivity criterion which corresponds to $N_S>3$ signal events.

\section{Results and discussion}

\begin{figure}[!ht]
    \centering
    \begin{minipage}[t]{0.49\textwidth}
        \centering
        \includegraphics[width=\linewidth]{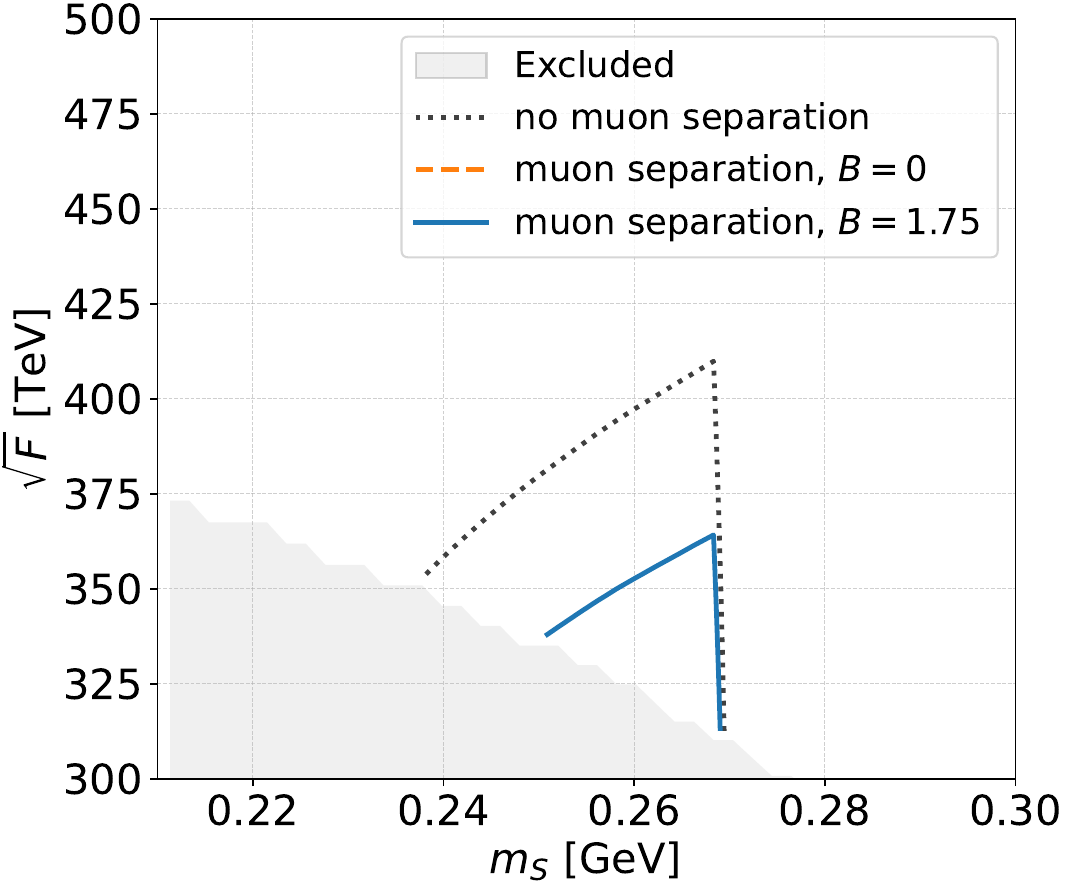}
        \par\small (a) Optimistic $\Delta_\mu = 1$ mm
    \end{minipage}\hfill
    \begin{minipage}[t]{0.49\textwidth}
        \centering
        \includegraphics[width=\linewidth]{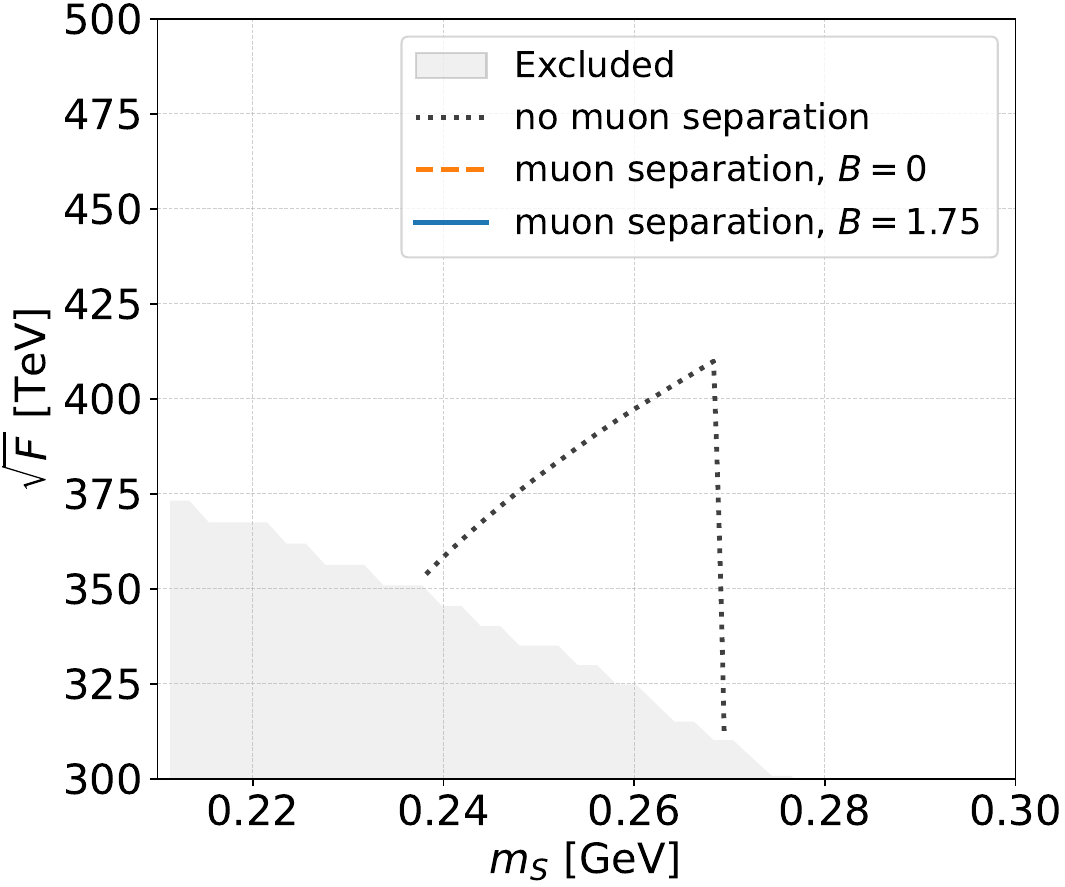}
        \par\small (b) Conservative $\Delta_\mu = 1$ cm
    \end{minipage}
    \caption{Sensitivity regions for $N_S>3$ (95\% CL) for the parameter set 1 as in \eqref{eq:sets} with $\tilde{m}^{LR} = 0$ for optimistic $\Delta_\mu = 1$ mm and conservative $\Delta_\mu = 1$ cm. Solid lines correspond to sgoldstinos decaying inside the detector with successful muon--antimuon separation. Dashed lines correspond to sgoldstinos decaying inside the detector with successful muon--antimuon separation in the absence of a magnetic field. Dotted lines correspond to sgoldstinos decaying into a muon--antimuon pair inside the detector. Gray areas correspond to the parameters excluded by the meson branching-fraction limits in Table~\ref{tab:Br_constraints}.}
    \label{fig:opt_sens}
\end{figure}

\begin{figure}[!ht]
    \centering
    \begin{minipage}[t]{0.49\textwidth}
        \centering
        \includegraphics[width=\linewidth]{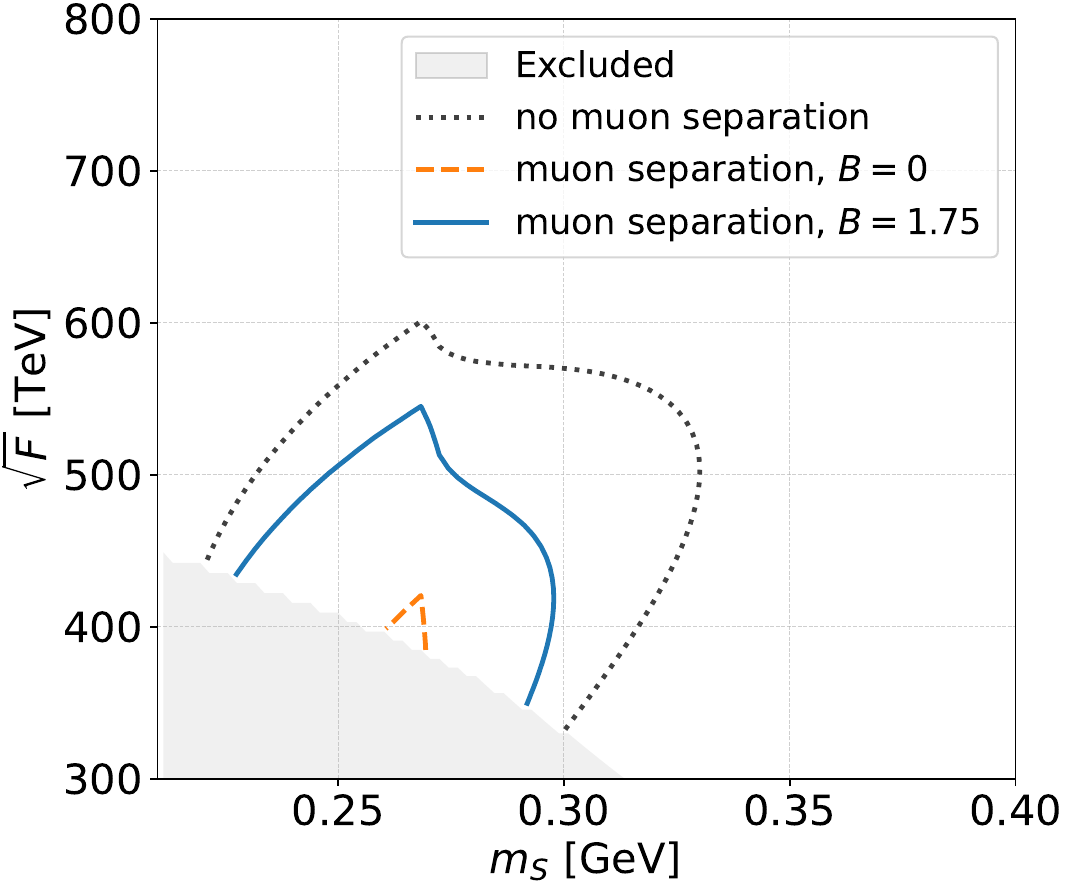}
        \par\small (a) Optimistic $\Delta_\mu = 1$ mm
    \end{minipage}\hfill
    \begin{minipage}[t]{0.49\textwidth}
        \centering
        \includegraphics[width=\linewidth]{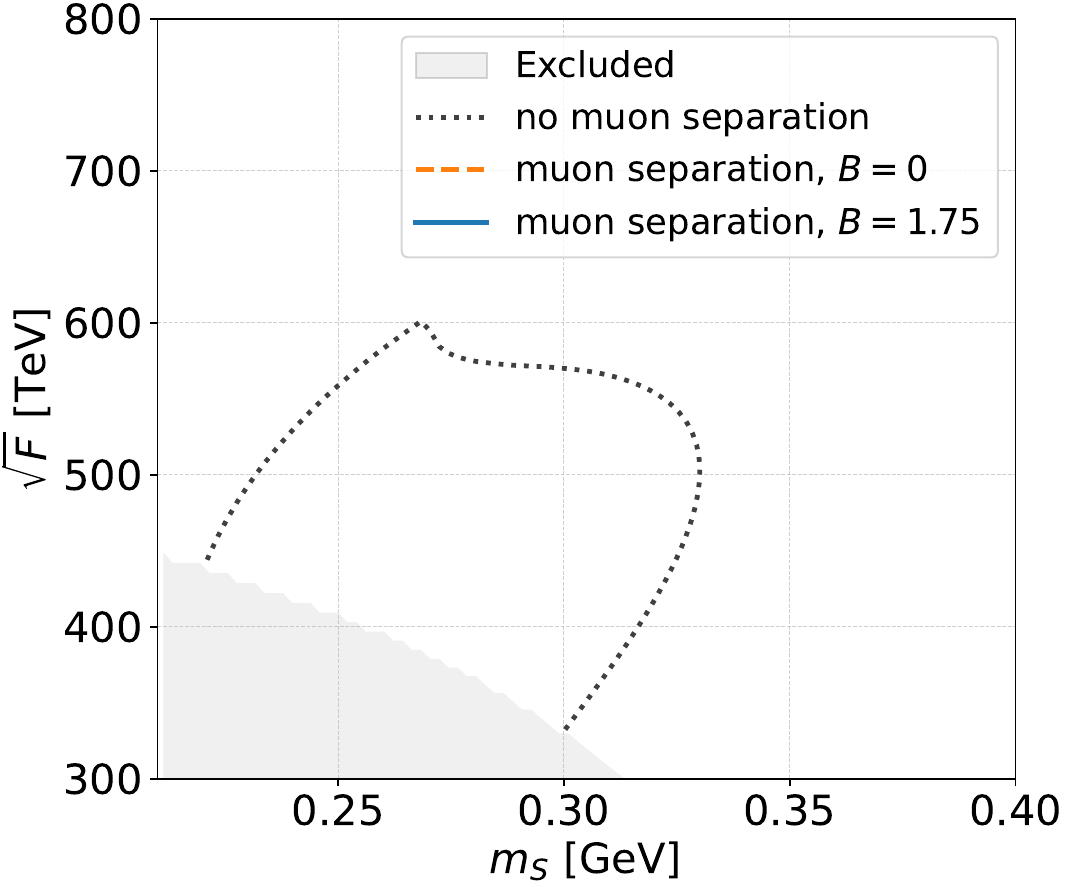}
        \par\small (b) Conservative $\Delta_\mu = 1$ cm
    \end{minipage}
    \caption{Sensitivity regions for $N_S>3$ (95\% CL) for the parameter set 2 as in \eqref{eq:sets} with $\tilde{m}^{LR} = 0$ for optimistic $\Delta_\mu = 1$ mm and conservative $\Delta_\mu = 1$ cm. Solid lines correspond to sgoldstinos decaying inside the detector with successful muon--antimuon separation. Dashed lines correspond to sgoldstinos decaying inside the detector with successful muon--antimuon separation in the absence of a magnetic field. Dotted lines correspond to sgoldstinos decaying into a muon--antimuon pair inside the detector. Gray areas correspond to the parameters excluded by the meson branching-fraction limits in Table~\ref{tab:Br_constraints}.}
    \label{fig:cons_sens}
\end{figure}

\begin{figure}[ht]
    \centering
    \begin{minipage}[t]{0.49\textwidth}
        \centering
        \includegraphics[width=\linewidth]{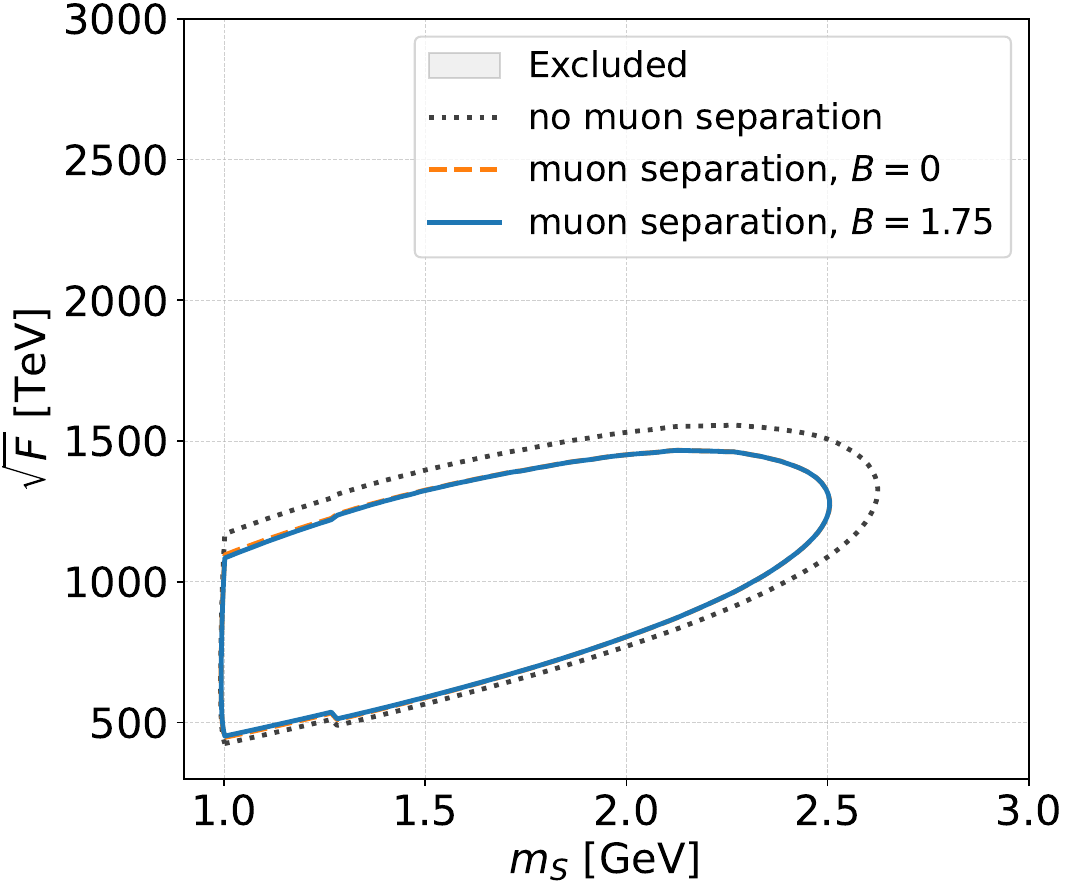}
        \par\small (a) Optimistic $\Delta_\mu = 1$ mm
    \end{minipage}\hfill
    \begin{minipage}[t]{0.49\textwidth}
        \centering
        \includegraphics[width=\linewidth]{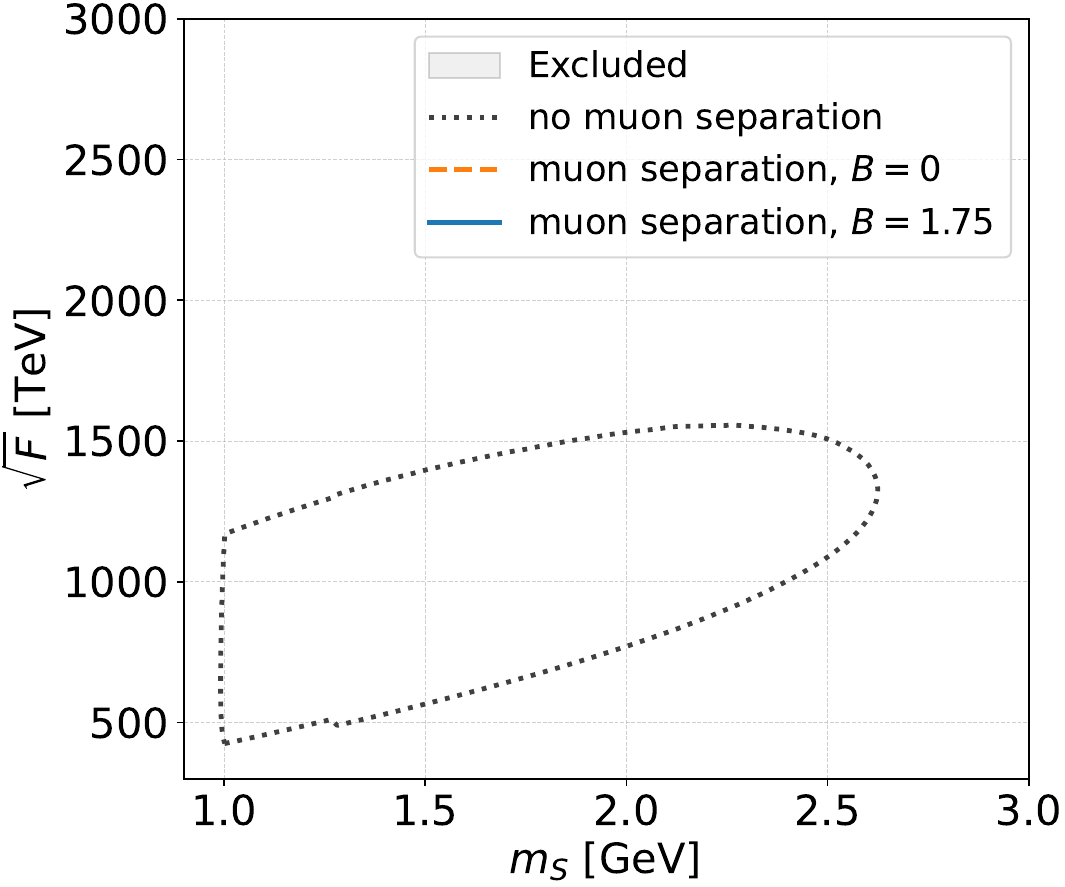}
        \par\small (b) Conservative $\Delta_\mu = 1$ cm
    \end{minipage}
    \caption{Sensitivity regions for $N_S>3$ (95\% CL) for the heavier-sgoldstino case with $\tilde{m}^{LR} = 0$ and set~2 parameters from Eq.~\eqref{eq:sets}. Solid lines correspond to sgoldstinos decaying inside the detector with successful muon--antimuon separation. Dashed lines correspond to sgoldstinos decaying inside the detector with successful muon--antimuon separation in the absence of a magnetic field. Dotted lines correspond to sgoldstinos decaying into a muon--antimuon pair inside the detector. Gray areas correspond to the parameters excluded by the meson branching-fraction limits in Table~\ref{tab:Br_constraints}.}
    \label{fig:sensitivity-B-heavy}
\end{figure}

The projected sensitivity for flavor-conserving couplings is shown in Fig.~\ref{fig:opt_sens} for the set 1 of model parameters and in Fig.~\ref{fig:cons_sens} for the set 2 of model parameters, see~\eqref{eq:sets}. The gray shaded regions correspond to the limits from meson branching fractions, see Table~\ref{tab:Br_constraints}. The dotted lines outline the parameter space in which more than 3 sgoldstinos reach the detector and decay into a muon--antimuon pair inside it. For the chosen parameter set and under the background-free assumption, this gives the maximal SND@HL-LHC parameter space available for the dimuon signature. The blue solid and orange dashed lines show the projected SND@HL-LHC sensitivity after applying the muon-separation condition described in Sec.~\ref{Muon_sep}. The blue solid lines correspond to the magnetized hadronic calorimeter, whereas the orange dashed lines correspond to the non-magnetized case. The loss of sensitivity at $m_S\approx 270$ MeV in panel (a) of Figs.~\ref{fig:opt_sens} and~\ref{fig:cons_sens} corresponds to the two-pion decay threshold. At the corresponding benchmark points, the mesonic decay widths are large and substantially reduce the sgoldstino lifetime. As a result, most sgoldstinos decay before reaching the detector. In set 2, shown in Figs.~\ref{fig:cons_sens}, the decay width into pions is smaller and is better matched to the SND detector position, 480 m downstream of the IP.

The comparison of the sensitivity contours in Figs.~\ref{fig:opt_sens} and~\ref{fig:cons_sens} shows that the muon--antimuon separation requirement has a substantial impact on the projected reach. In the flavor-conserving scenarios considered here, the dominant contribution to the signal comes from $B$-meson decays. As shown in panel~(b) of Fig.~\ref{fig:Sspectra}, most of the corresponding sgoldstinos that decay inside the detector have momenta of approximately $700$--$900$ GeV. Consequently, the produced dimuon pair is strongly collimated, causing the conservative separation criteria, $\Delta_\mu=1$ cm, to reject almost all of these decays, see Fig.~\ref{fig:muon_factor}. For the optimistic criterion, $\Delta_\mu=1$ mm, the effective momentum cutoff is higher and a larger fraction of the sgoldstino decays produces resolvable muon and antimuon tracks. Nevertheless, the comparison between the solid and dashed contours demonstrates that a high spatial resolution alone is not sufficient. %The charge-dependent bending in the magnetized HCAL can substantially increase the accepted event rate and restore sensitivity in parts of the parameter space.
The additional muon tracks separation in the magnetic field of HCAL can substantially increase the accepted number of events and restore sensitivity.

The sgoldstino spectrum from kaon decays is softer, as shown in panel~(a) of Fig.~\ref{fig:Sspectra}, and therefore less affected by the muon-separation requirement. However, the relatively long lifetimes and large momenta of the $K_L$ and $K^{\pm}$ make the probability of their decaying before reaching the structural elements very small, while the $K_S$ contribution is suppressed as discussed above. The resulting kaon contribution is therefore negligible. %despite its more favorable dimuon-separation efficiency.

A separate case arises for heavier sgoldstinos whose hadronic decay width is described by the gluonic channel; see Fig.~\ref{fig:sensitivity-B-heavy}. In this region, in the absence of flavor-violating couplings, the only viable source of sgoldstinos is $B$-meson decay. For a fixed momentum, the larger sgoldstino mass corresponds to a smaller Lorentz boost and hence a larger intrinsic opening angle of the muon--antimuon pair, making the two tracks easier to separate. This effect is visible in Fig.~\ref{fig:sensitivity-B-heavy}. For the conservative criterion, $\Delta_\mu=1$ cm, the intrinsic separation and the additional magnetic bending are still insufficient to satisfy separation requirement. Conversely, for the optimistic criterion, $\Delta_\mu=1$ mm, almost all sgoldstino decays inside the detector satisfy the separation requirement. In this case, the spatial resolution alone is often sufficient, and the magnetic field effect is negligible (in Fig.~\ref{fig:sensitivity-B-heavy}(a) the dashed line is almost coincides with the solid one). The magnitude of this effect is nevertheless strongly model-dependent because it also depends on the sgoldstino lifetime. For a fixed detector position, longer lifetimes can allow lower-momentum particles to reach the detector, improving the dimuon-separation efficiency. This improvement, however, is not independent of the production rate: in many models the same diminishing of couplings that increases the lifetime also suppresses the production rate of the new particle. Therefore, models in which meson decays produce new particles are not automatically limited by the muon-separation factor. The role of the magnetic field should thus be interpreted as conditional, depending on the assumed spatial-separation threshold, parent-meson kinematics, particle lifetime, and production rate rather than as a universal requirement for all sgoldstino signals.

\begin{figure}[ht]
    \centering
    \begin{minipage}[t]{0.49\textwidth}
        \centering
         \includegraphics[width=\linewidth]{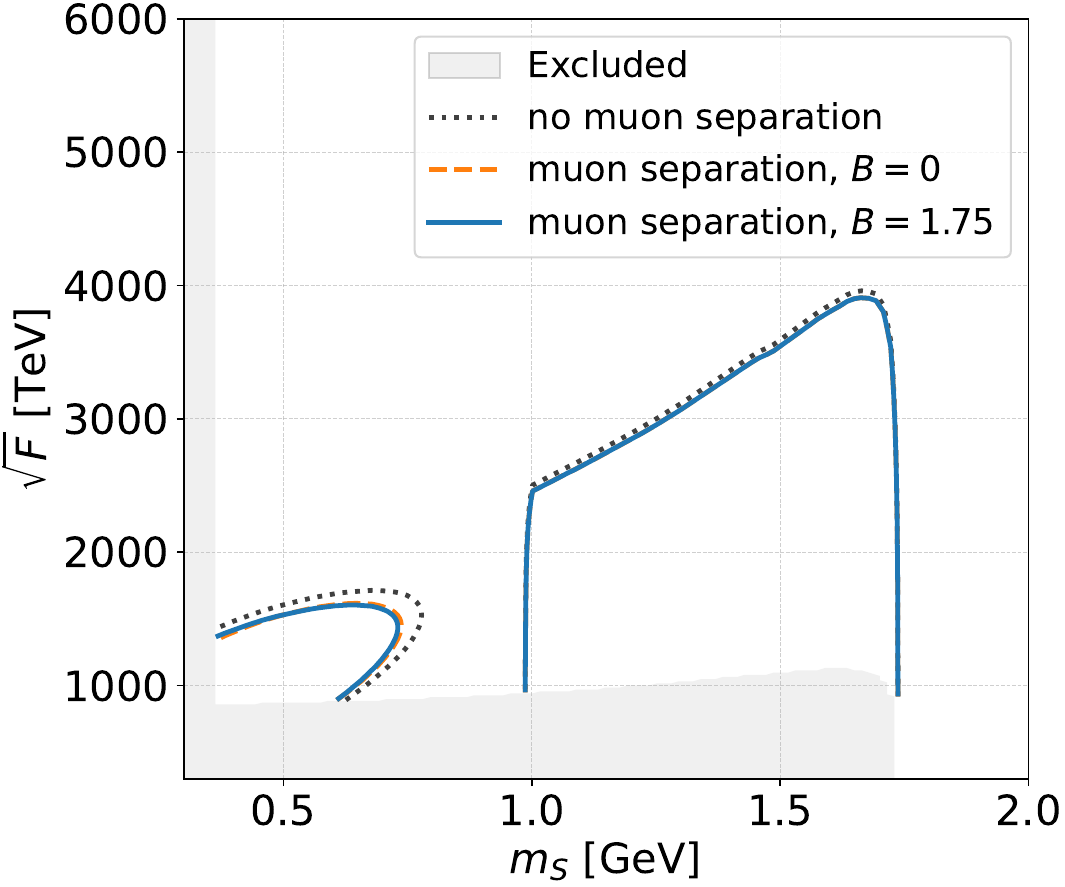}
        \par\small (a) Optimistic $\Delta_\mu = 1$ mm
    \end{minipage}\hfill
    \begin{minipage}[t]{0.49\textwidth}
        \centering
        \includegraphics[width=\linewidth]{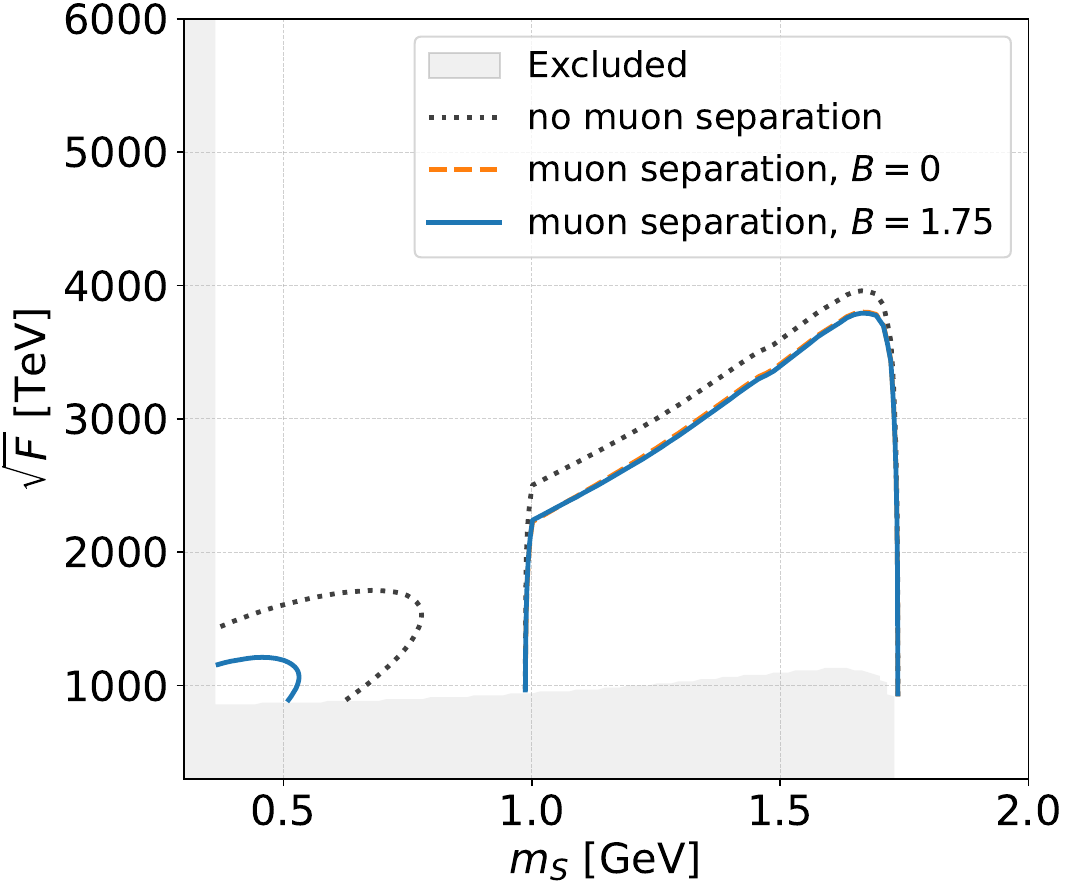}
        \par\small (b) Conservative $\Delta_\mu = 1$ cm
    \end{minipage}
    \caption{Sensitivity regions for $N_S>3$ (95\% CL) with a nonzero flavor-violating contribution, $\tilde{m}^{LR}=30$ GeV. Solid lines correspond to sgoldstinos decaying inside the detector with successful muon--antimuon separation. Dashed lines correspond to sgoldstinos decaying inside the detector with successful muon--antimuon separation in the absence of a magnetic field. Dotted lines correspond to sgoldstinos decaying into a muon--antimuon pair inside the detector. Model parameters correspond to set 2 in Eq.~\eqref{eq:sets}. Gray areas correspond to the parameters excluded by the meson branching-fraction limits in Table~\ref{tab:Br_constraints}.}
    \label{fig:sensitivity-fv}
\end{figure}

\begin{figure}[!ht]  
    \centering
    \begin{minipage}[t]{0.49\textwidth}
        \centering
        \includegraphics[width=\linewidth]{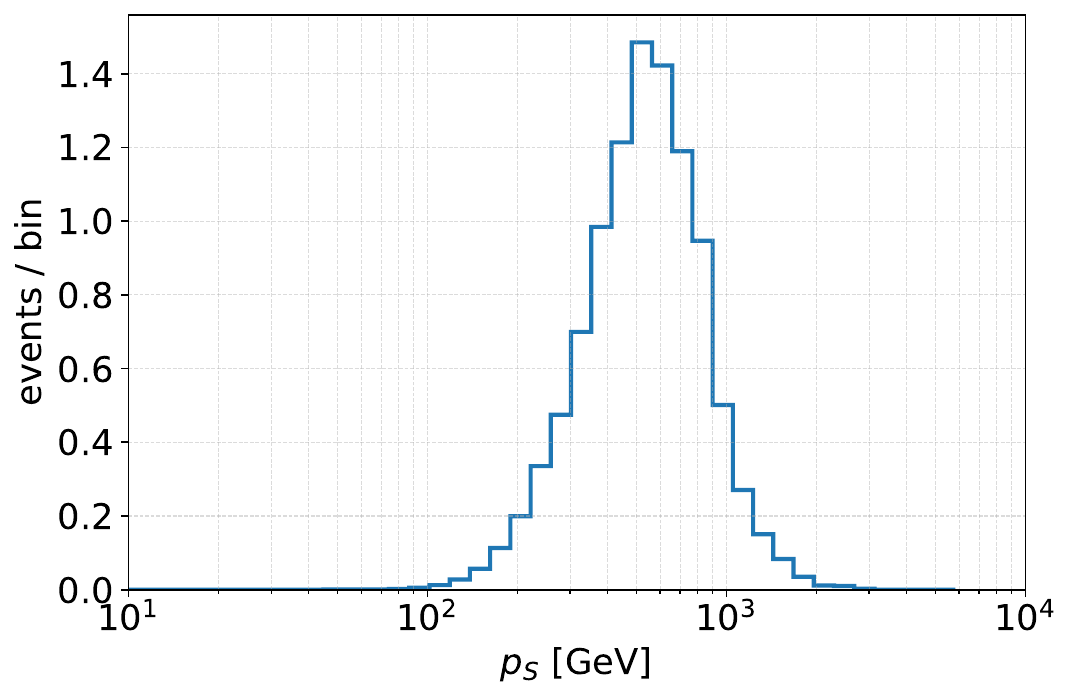}
        \par\small (a) $D$ channel. $m_S=0.5$ GeV, $\sqrt{F}=1000$ TeV.
    \end{minipage}\hfill
    \begin{minipage}[t]{0.49\textwidth}
        \centering
        \includegraphics[width=\linewidth]{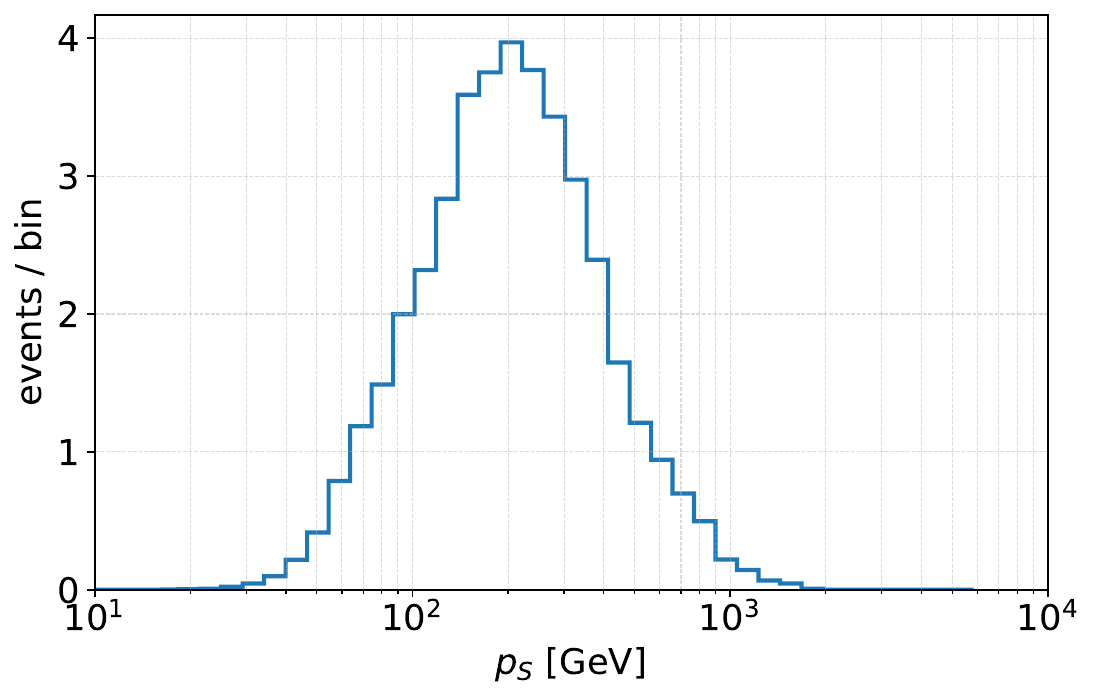}
        \par\small (b) $D$ channel. $m_S=1.5$ GeV, $\sqrt{F}=2000$ TeV.
    \end{minipage}
\caption{Sgoldstino momentum distributions for decays inside the detector volume. Set~2 parameters from Eq.~\eqref{eq:sets} and $\tilde{m}^{LR}=30$ GeV are used.}
\label{fig:D_fv_spectra}
\end{figure}

We now turn to the case with flavor-violating couplings and set $\tilde{m}^{LR}=30$ GeV. This opens a new mass range with an enhanced branching fraction for $D$-meson decays to sgoldstino. The corresponding sensitivity is shown in Fig.~\ref{fig:sensitivity-fv}, where only $D$-meson decays are considered as a source of sgoldstinos. In this case, models with $m_S<m_K-m_\pi$ are constrained by limits on kaon decay branching fractions, and $B$-meson decays induced by the flavor-violating coupling do not introduce additional sensitivity regions. Therefore, among the two model parameter sets SND is sensitive to sgoldstinos only for set 2, with $M_{\gamma \gamma}= 1$ TeV and $M_3=3$ TeV. For the lighter sgoldstino, the sensitivity region is centered near $\sqrt{F}=1000$ TeV, compared to $\sqrt{F}=2000$ TeV for the heavier sgoldstino. This difference translates into different sgoldstino lifetimes and different momentum spectra for sgoldstinos decaying inside the detector; see Fig.~\ref{fig:D_fv_spectra}. The lighter-sgoldstino distribution has a larger mean momentum, $p_S \approx 600$ GeV, compared to $p_S \approx 200$ GeV for the heavier sgoldstino. This explains the pronounced change in sensitivity. At the benchmark points shown in Fig.~\ref{fig:D_fv_spectra} heavier sgoldstinos have longer lifetimes and smaller mean momenta. Therefore, these sgoldstinos are accepted by the muon--antimuon separation factor. Unlike $B$-meson decays, flavor-violating $D$-meson decays produce much more sgoldstinos with smaller momenta; therefore, they do not suffer a large sensitivity loss in the conservative case. For lighter sgoldstinos, the situation is somewhat different. In this region, sgoldstinos have larger mean momentum and are cut off by the muon-separation factor in the conservative case. However, the magnetic field raises the momentum cutoff enough to accept additional sgoldstino decays. Because of the large overall number of sgoldstinos produced in $D$-meson decays in this parameter set, this opens a sensitivity region for the conservative muon--antimuon separation condition.

In addition to the scalar sgoldstino considered in this work, the same supersymmetry-breaking sector also contains a pseudoscalar sgoldstino. Its phenomenology is expected to be close to the scalar case except in the intermediate hadronic region, in particular for $2m_\pi < m_{S(P)}<1$ GeV~\cite{Gorbunov:2000th,Astapov:2015otc}. In this region, the pseudoscalar nature of this state changes the hadronic decay pattern: the two-pion channel is forbidden, and the lowest pion final state is instead the three-pion channel. As a result, for the same set of model parameters, the mass interval in which the dimuon decay mode dominates can extend up to approximately the three-pion threshold, $m_P \simeq 3m_\pi$. The mesonic decay width is generally expected to be smaller than that for the scalar sgoldstino, and the same applies to the production rate in meson decays~\cite{Demidov:2022fas}. Including pseudoscalar sgoldstinos may therefore somewhat extend the SND@HL-LHC sensitivity in the region $2m_\pi < m_P < 3m_\pi$. Nevertheless, from the point of view of muon-track separation, it is not expected to introduce qualitatively new effects compared with the scalar case.

\section{Conclusion}
We have investigated the prospects for scalar-sgoldstino searches at the proposed SND@HL-LHC detector. For the representative supersymmetry-breaking scenarios considered in this work, SND@HL-LHC can probe sgoldstino masses from the dimuon threshold, $m_S \simeq 2m_\mu$, up to $m_S \approx 2.5$ GeV and supersymmetry-breaking scales as large as $\sqrt{F}\approx 4000$ TeV. The exact sensitivity depends on the sgoldstino couplings, lifetime, and dominant meson-production channel.

We have also quantified the impact of the muon--antimuon separation criteria on the projected detector sensitivity. This requirement can substantially reduce the number of observable events, particularly for highly boosted light sgoldstinos producing strongly collimated muon pairs. The separation efficiency is governed primarily by the spatial resolution of the HCAL. For light sgoldstinos, the magnetic field plays a crucial role and can significantly extend the sensitivity region. For heavier sgoldstinos, however, the decay kinematics typically lead to a larger intrinsic opening angle between the muons. In this case, sufficiently high spatial resolution alone is often enough to resolve the two tracks, and the magnetic field provides only a moderate improvement.

\section{Acknowledgments}
The authors thank Sergei Demidov, Dmitry Gorbunov, and Mikhail Vysotsky for helpful discussions and valuable insights. The work of DK was supported by the Russian Science Foundation under grant No.~25-12-00309.

\bibliographystyle{unsrtnat}

\bibliography{refs}
\end{document}